%% file: main.tex
\documentclass[twocolumn,tighten,times,twocolappendix]{aastex631}

\usepackage{amsmath}
\usepackage{xcolor}

\usepackage{booktabs}

\graphicspath{{images/}}

\newcommand{\psedit}{}

\newcommand{\mr}{\mathrm}
\newcommand{\pr}{\Phi(r)}
\newcommand{\prp}{\Phi(r^\prime/\sqrt{K})}
\newcommand{\prk}{\Phi(r/\sqrt{K})}
\newcommand{\aver}{\langle r \rangle}
\newcommand{\averp}{\langle r^\prime \rangle}

\usepackage{hyperref}
\usepackage{graphicx} 

\shortauthors{Lin et al.}

\begin{document}

\title{Do All Fast Radio Bursts Repeat? Constraints from CHIME/FRB Far Side-Lobe FRBs}

\correspondingauthor{Paul Scholz}
\email{pscholz@yorku.ca}

\author[0000-0001-7453-4273]{Hsiu-Hsien Lin}
\affiliation{Institute of Astronomy and Astrophysics, Academia Sinica, Astronomy-Mathematics Building, No. 1, Sec. 4, Roosevelt Road, Taipei 10617, Taiwan}
  \affiliation{Canadian Institute for Theoretical Astrophysics, 60 St.~George Street, Toronto, ON M5S 3H8, Canada}
\author[0000-0002-7374-7119]{Paul Scholz}
  \affiliation{Dunlap Institute for Astronomy \& Astrophysics, University of Toronto, 50 St.~George Street, Toronto, ON M5S 3H4, Canada}
  \affiliation{Department of Physics and Astronomy, York University, 4700 Keele Street, Toronto, ON MJ3 1P3, Canada}
\author[0000-0002-3616-5160]{Cherry Ng}
  \affiliation{Dunlap Institute for Astronomy \& Astrophysics, University of Toronto, 50 St.~George Street, Toronto, ON M5S 3H4, Canada}
  \affiliation{Laboratoire de Physique et Chimie de l’Environnement et de l’Espace, Université d’Orléans / CNRS, 45071 Orléans Cedex 02, France}
\author[0000-0003-2155-9578]{Ue-Li Pen}
  \affiliation{Institute of Astronomy and Astrophysics, Academia Sinica, Astronomy-Mathematics Building, No. 1, Sec. 4, Roosevelt Road, Taipei 10617, Taiwan}
  \affiliation{Canadian Institute for Theoretical Astrophysics, 60 St.~George Street, Toronto, ON M5S 3H8, Canada}
  \affiliation{Canadian Institute for Advanced Research, MaRS Centre, West Tower, 661 University Avenue, Suite 505, Toronto, ON M5G 1M1, Canada}
  \affiliation{Dunlap Institute for Astronomy \& Astrophysics, University of Toronto, 50 St.~George Street, Toronto, ON M5S 3H4, Canada}
  \affiliation{Perimeter Institute for Theoretical Physics, 31 Caroline Street N, Waterloo, ON N25 2YL, Canada}
\author[0000-0002-3615-3514]{Mohit Bhardwaj}
  \affiliation{Department of Physics, Carnegie Mellon University, 5000 Forbes Avenue, Pittsburgh, 15213, PA, USA}
\author[0000-0002-3426-7606]{Pragya Chawla}
  \affiliation{Anton Pannekoek Institute for Astronomy, University of Amsterdam, Science Park 904, 1098 XH Amsterdam, The Netherlands}
\author[0000-0002-8376-1563]{Alice P.~Curtin}
  \affiliation{Trottier Space Institute, McGill University, 3550 rue University, Montr\'eal, QC H3A 2A7, Canada}
  \affiliation{Department of Physics, McGill University, 3600 rue University, Montr\'eal, QC H3A 2T8, Canada}
\author[0000-0001-7931-0607]{Dongzi Li}
  \affiliation{Cahill Center for Astronomy and Astrophysics, California Institute of Technology, 1216 E California Boulevard, Pasadena, CA 91125, USA}
\author[0000-0002-7333-5552]{Laura Newburgh}
  \affiliation{Department of Physics, Yale University, New Haven, CT 06520, USA}
\author[0000-0001-6967-7253]{Alex Reda}
  \affiliation{Department of Physics, Yale University, New Haven, CT 06520, USA}
\author[0000-0003-3154-3676]{Ketan R.~Sand}
  \affiliation{Trottier Space Institute, McGill University, 3550 rue University, Montr\'eal, QC H3A 2A7, Canada}
  \affiliation{Department of Physics, McGill University, 3600 rue University, Montr\'eal, QC H3A 2T8, Canada}
\author[0000-0003-2548-2926]{Shriharsh P.~Tendulkar}
  \affiliation{Department of Astronomy and Astrophysics, Tata Institute of Fundamental Research, Mumbai, 400005, India}
  \affiliation{National Centre for Radio Astrophysics, Post Bag 3, Ganeshkhind, Pune, 411007, India}
\author[0000-0001-5908-3152]{Bridget Andersen}
  \affiliation{Department of Physics, McGill University, 3600 rue University, Montr\'eal, QC H3A 2T8, Canada}
  \affiliation{Trottier Space Institute, McGill University, 3550 rue University, Montr\'eal, QC H3A 2A7, Canada}
\author[0000-0003-3772-2798]{Kevin Bandura}
  \affiliation{Lane Department of Computer Science and Electrical Engineering, 1220 Evansdale Drive, PO Box 6109 Morgantown, WV 26506, USA}
  \affiliation{Center for Gravitational Waves and Cosmology, West Virginia University, Chestnut Ridge Research Building, Morgantown, WV 26505, USA}
\author[0000-0002-1800-8233]{Charanjot Brar}
  \affiliation{Department of Physics, McGill University, 3600 rue University, Montr\'eal, QC H3A 2T8, Canada}
  \affiliation{Trottier Space Institute, McGill University, 3550 rue University, Montr\'eal, QC H3A 2A7, Canada}
\author[0000-0003-2047-5276]{Tomas Cassanelli}
  \affiliation{Department of Electrical Engineering, Universidad de Chile, Av. Tupper 2007, Santiago 8370451, Chile}  
\author[0000-0001-6422-8125]{Amanda M.~Cook}
  \affiliation{Dunlap Institute for Astronomy \& Astrophysics, University of Toronto, 50 St.~George Street, Toronto, ON M5S 3H4, Canada}
  \affiliation{David A.~Dunlap Department of Astronomy \& Astrophysics, University of Toronto, 50 St.~George Street, Toronto, ON M5S 3H4, Canada}
\author[0000-0001-7166-6422]{Matt Dobbs}
  \affiliation{Department of Physics, McGill University, 3600 rue University, Montr\'eal, QC H3A 2T8, Canada}
  \affiliation{Trottier Space Institute, McGill University, 3550 rue University, Montr\'eal, QC H3A 2A7, Canada} 
\author[0000-0003-4098-5222]{Fengqiu Adam Dong}
  \affiliation{Department of Physics and Astronomy, University of British Columbia, 6224 Agricultural Road, Vancouver, BC V6T 1Z1 Canada}
\author[0000-0003-3734-8177]{Gwendolyn Eadie}
  \affiliation{David A.~Dunlap Department of Astronomy \& Astrophysics, University of Toronto, 50 St.~George Street, Toronto, ON M5S 3H4, Canada}
  \affiliation{Department of Statistical Sciences, University of Toronto, Toronto, ON M5S 3G3, Canada}  
\author[0000-0001-8384-5049]{Emmanuel Fonseca}
  \affiliation{Department of Physics and Astronomy, West Virginia University, P.O. Box 6315, Morgantown, WV 26506, USA }
  \affiliation{Center for Gravitational Waves and Cosmology, West Virginia University, Chestnut Ridge Research Building, Morgantown, WV 26505, USA}
\author[0000-0002-3382-9558]{B.~M.~Gaensler}
  \affiliation{Dunlap Institute for Astronomy \& Astrophysics, University of Toronto, 50 St.~George Street, Toronto, ON M5S 3H4, Canada}
  \affiliation{David A.~Dunlap Department of Astronomy \& Astrophysics, University of Toronto, 50 St.~George Street, Toronto, ON M5S 3H4, Canada}
\author[0000-0001-5553-9167]{Utkarsh Giri}
  \affiliation{Department of Physics, University of Wisconsin-Madison, 1150 University Ave, Madison, WI 53706, USA}
\author[0000-0002-3654-4662]{Antonio Herrera-Martin}
  \affiliation{David A.~Dunlap Department of Astronomy \& Astrophysics, University of Toronto, 50 St.~George Street, Toronto, ON M5S 3H4, Canada}
\author[0000-0001-7301-5666]{Alex S. Hill}
  \affiliation{Department of Computer Science, Math, Physics, \& Statistics, University of British Columbia, Okanagan Campus, Kelowna, BC V1V 1V7, Canada}
  \affiliation{Dominion Radio Astrophysical Observatory, Herzberg Research Centre for Astronomy and Astrophysics, National Research Council Canada, PO Box 248, Penticton, BC V2A 6J9, Canada}
\author[0000-0003-4810-7803]{Jane Kaczmarek}
  \affiliation{Dominion Radio Astrophysical Observatory, Herzberg Research Centre for Astronomy and Astrophysics, National Research Council Canada, PO Box 248, Penticton, BC V2A 6J9, Canada}
  \affiliation{CSIRO Space \& Astronomy, Parkes Observatory, P.O. Box 276, Parkes NSW 2870, Australia}
\author[0000-0002-3354-3859]{Joseph	Kania}
  \affiliation{Department of Physics and Astronomy, West Virginia University, P.O. Box 6315, Morgantown, WV 26506, USA }
  \affiliation{Center for Gravitational Waves and Cosmology, West Virginia University, Chestnut Ridge Research Building, Morgantown, WV 26505, USA}
\author[0000-0001-9345-0307]{Victoria Kaspi}
  \affiliation{Department of Physics, McGill University, 3600 rue University, Montr\'eal, QC H3A 2T8, Canada}
  \affiliation{Trottier Space Institute, McGill University, 3550 rue University, Montr\'eal, QC H3A 2A7, Canada}
\author[0009-0005-7115-3447]{Kholoud Khairy}
  \affiliation{Lane Department of Computer Science and Electrical Engineering, 1220 Evansdale Drive, PO Box 6109  Morgantown, WV 26506, USA}
  \affiliation{Center for Gravitational Waves and Cosmology, West Virginia University, Chestnut Ridge Research Building, Morgantown, WV 26505, USA}
\author[0000-0003-2116-3573]{Adam E.~Lanman}
  \affiliation{Trottier Space Institute, McGill University, 3550 rue University, Montr\'eal, QC H3A 2A7, Canada}
  \affiliation{Department of Physics, McGill University, 3600 rue University, Montr\'eal, QC H3A 2T8, Canada}
\author[0000-0002-4209-7408]{Calvin Leung}
  \affiliation{MIT Kavli Institute for Astrophysics and Space Research, Massachusetts Institute of Technology, 77 Massachusetts Ave, Cambridge, MA 02139, USA}
  \affiliation{Department of Physics, Massachusetts Institute of Technology, 77 Massachusetts Ave, Cambridge, MA 02139, USA}
\affiliation{NHFP Einstein Fellow}  
\author[0000-0002-4279-6946]{Kiyoshi W.~Masui}
  \affiliation{MIT Kavli Institute for Astrophysics and Space Research, Massachusetts Institute of Technology, 77 Massachusetts Ave, Cambridge, MA 02139, USA}
  \affiliation{Department of Physics, Massachusetts Institute of Technology, 77 Massachusetts Ave, Cambridge, MA 02139, USA}
\author[0000-0002-0772-9326]{Juan Mena-Parra}
  \affiliation{Dunlap Institute for Astronomy \& Astrophysics, University of Toronto, 50 St.~George Street, Toronto, ON M5S 3H4, Canada}
  \affiliation{David A.~Dunlap Department of Astronomy \& Astrophysics, University of Toronto, 50 St.~George Street, Toronto, ON M5S 3H4, Canada}  
\author[0000-0001-8845-1225]{Bradley W.	Meyers}
  \affiliation{International Centre for Radio Astronomy Research (ICRAR), Curtin University, Bentley WA 6102 Australia}
\author[0000-0002-2551-7554]{Daniele Michilli}
  \affiliation{MIT Kavli Institute for Astrophysics and Space Research, Massachusetts Institute of Technology, 77 Massachusetts Ave, Cambridge, MA 02139, USA}
  \affiliation{Department of Physics, Massachusetts Institute of Technology, 77 Massachusetts Ave, Cambridge, MA 02139, USA}
\author[0000-0001-8292-0051]{Nikola Milutinovic}
  \affiliation{Department of Physics and Astronomy, University of British Columbia, 6224 Agricultural Road, Vancouver, BC V6T 1Z1 Canada}
\author[0000-0002-2465-8937]{Anna Ordog}
  \affiliation{Dominion Radio Astrophysical Observatory, Herzberg Research Centre for Astronomy and Astrophysics, National Research Council Canada, PO Box 248, Penticton, BC V2A 6J9, Canada}
  \affiliation{Department of Computer Science, Math, Physics, \& Statistics, University of British Columbia, Okanagan Campus, Kelowna, BC V1V 1V7, Canada}
\author[0000-0002-8912-0732]{Aaron B.~Pearlman}
  \affiliation{Department of Physics, McGill University, 3600 rue University, Montr\'eal, QC H3A 2T8, Canada}
  \affiliation{Trottier Space Institute, McGill University, 3550 rue University, Montr\'eal, QC H3A 2A7, Canada}
  \affiliation{Banting Fellow}
  \affiliation{McGill Space Institute Fellow}
  \affiliation{FRQNT Postdoctoral Fellow}
\author[0000-0002-4795-697X]{Ziggy Pleunis}
  \affiliation{Dunlap Institute for Astronomy \& Astrophysics, University of Toronto, 50 St.~George Street, Toronto, ON M5S 3H4, Canada}
\author[0000-0001-7694-6650]{Masoud Rafiei-Ravandi}
  \affiliation{Department of Physics, McGill University, 3600 rue University, Montr\'eal, QC H3A 2T8, Canada}
  \affiliation{Trottier Space Institute, McGill University, 3550 rue University, Montr\'eal, QC H3A 2A7, Canada}
\author[0000-0003-1842-6096]{Mubdi Rahman}
  \affiliation{Sidrat Research, 124 Merton Street, Suite 507, Toronto, ON M4S 2Z2, Canada}    
\author[0000-0001-5799-9714]{Scott Ransom}
  \affiliation{National Radio Astronomy Observatory, 520 Edgemont Rd, Charlottesville, VA 22903, USA}
\author[0000-0001-5504-229X]{Pranav Sanghavi}
  \affiliation{Department of Physics, Yale University, New Haven, CT 06520, USA}
\author[0000-0002-6823-2073]{Kaitlyn Shin}
  \affiliation{MIT Kavli Institute for Astrophysics and Space Research, Massachusetts Institute of Technology, 77 Massachusetts Ave, Cambridge, MA 02139, USA}
  \affiliation{Department of Physics, Massachusetts Institute of Technology, 77 Massachusetts Ave, Cambridge, MA 02139, USA}  
\author[0000-0002-2088-3125]{Kendrick Smith}
  \affiliation{Perimeter Institute for Theoretical Physics, 31 Caroline Street N, Waterloo, ON N25 2YL, Canada}
\author[0000-0001-9784-8670]{Ingrid Stairs}
  \affiliation{Department of Physics and Astronomy, University of British Columbia, 6224 Agricultural Road, Vancouver, BC V6T 1Z1 Canada}
\author[0000-0002-9761-4353]{David C Stenning}
  \affiliation{Department of Statistics \& Actuarial Science, Simon Fraser University, Burnaby, BC, Canada}
\author[0000-0003-4535-9378]{Keith Vanderlinde}
  \affiliation{Dunlap Institute for Astronomy \& Astrophysics, University of Toronto, 50 St.~George Street, Toronto, ON M5S 3H4, Canada}
  \affiliation{David A.~Dunlap Department of Astronomy \& Astrophysics, University of Toronto, 50 St.~George Street, Toronto, ON M5S 3H4, Canada}
\author[0000-0001-7314-9496]{Dallas Wulf}
  \affiliation{Department of Physics, McGill University, 3600 rue University, Montr\'eal, QC H3A 2T8, Canada}
  \affiliation{Trottier Space Institute, McGill University, 3550 rue University, Montr\'eal, QC H3A 2A7, Canada}  

\begin{abstract}
  We report ten fast radio bursts (FRBs) detected in the far side-lobe region (i.e., $\geq 5^\circ$ off-meridian) of the Canadian Hydrogen Intensity Mapping Experiment (CHIME) from 2018 August 28 to 2021 August 31. We localize the bursts by fitting their spectra with a model of the CHIME/FRB synthesized beam response. \psedit{We find that the far side-lobe events have on average $\sim$500 times greater fluxes than events detected in CHIME's main lobe. We show that the side-lobe sample is therefore statistically $\sim$20 times closer than the main-lobe sample. We find promising host galaxy candidates (P$_{\rm cc} < 1$\%) for two of the FRBs, 20190112B and 20210310B, at distances of 38 and 16 Mpc, respectively.}  CHIME/FRB did not observe repetition of similar brightness from the uniform sample of 10 side-lobe FRBs in a total exposure time of 35580 hours. Under the assumption of Poisson-distributed bursts, we infer that the mean repetition interval above the \psedit{detection} threshold of the far side-lobe events is longer than 11880 hours, which is at least 2380 times larger than the interval from known CHIME/FRB detected repeating sources, with some caveats, notably that very narrow-band events could have been missed. Our results from these far side-lobe events suggest one of two scenarios: either (1) all FRBs repeat and the repetition intervals span a wide range, with high-rate repeaters being a rare subpopulation, or (2) non-repeating FRBs are a distinct population different from known repeaters. 
\end{abstract}


\section{Introduction} \label{sec:intro}

Fast radio bursts (FRBs) are bright radio transients with milliseconds duration, cosmological origin, and unknown physical mechanism \citep{2007Sci...318..777L, 2019A&ARv..27....4P, 2019ARA&A..57..417C}. In the past decade, over 600 FRBs have been published \citep{ 2021ApJS..257...59A}, of which fifty have been seen to repeat \citep{2016Natur.531..202S, 2019Natur.566..235C, 2019ApJ...885L..24C, 2020ApJ...891L...6F, 2020Natur.582..351C, 2023arXiv230108762T}. Nearly two dozen FRBs have been localized to their host galaxies using interferometry. With galaxy identification, redshift and host type can be determined, which are crucial for understanding the nature of FRBs \citep{2017Natur.541...58C, 2019Natur.572..352R, 2020Natur.577..190M, 2020Natur.581..391M, 2022Natur.602..585K}. 

There is a diversity of physical models for FRBs \citep[see][]{2019PhR...821....1P} with many models allowing for repetition and many not. One possibility is that some FRBs do not repeat, which motivates cataclysmic scenarios such as a merger system for black holes or neutron stars \citep{2013PASJ...65L..12T, 2015ApJ...814L..20M}. Another possibility is that all FRBs will be seen to repeat as long as the observation time is long enough and with sufficient instrumental sensitivity. Two repeaters have been observed to show periodic activity windows \citep{2020Natur.582..351C, 2020MNRAS.495.3551R, 2021MNRAS.500..448C}. For some repeaters, the bursts appear clustered, and the waiting times could be from a few hours to several months \citep{2021Natur.598..267L, 2022ApJ...927...59L}. In addition, there is an observed dichotomy between the morphology of apparent non-repeaters and repeater bursts \citep{2021ApJ...923....1P}.

To detect an FRB, either the telescope must have a high sensitivity that allows it to detect apparently faint, more common events, or a long exposure time, so that it can detect apparently bright, rare events. For a radio telescope, most of the sensitivity is directed towards a ``main lobe'', where the telescope is pointed (along the meridian in the case of CHIME), but the telescope and its synthesized beams have lower sensitivity out to the horizon in ``sidelobes''. With sufficient exposure time, events will be detected in the sidelobes. Since sidelobes are less sensitive than the main lobe, events detected in sidelobes will be rare and apparently bright. Assuming that all FRBs have the same luminosity function, apparently bright events are typically closer than faint events, because nearby sources tend to be brighter than farther sources \citep{2022PASP..134i4106L}.

For instance, CHIME/FRB detected apparently bright bursts from the Galactic magnetar SGR 1935+2154 in the far side-lobe regions \citep{2020Natur.587...54C, 2022ATel15681....1D}. Moreover, the Parkes (Murriyang) telescope detected another apparently bright and low-DM FRB, FRB 20110214A, in the sidelobes of two of the beams of the multi-beam receiver \citep{2019MNRAS.482.3109P}. With $\sim$60 hrs of follow-up observations, the Parkes telescope did not detect any repeating bursts \citep{2019MNRAS.482.3109P}. The non-detection of repetition raises a key question in the FRB field: \textit{Do all FRBs repeat?} And if they do, \textit{how long an exposure time is needed to observe the repetition from an apparent non-repeater?} \psedit{Events detected in CHIME/FRB far side lobes have daily exposure times of several hours at a sensitivity threshold of the initial detection. They thus probe a regime that is different from that in the main lobe sample: extremely long per-source exposure times for the detection of bright bursts. These long exposures are therefore useful to answer this question}. 


In this paper, we present 10 far side-lobe FRBs detected by CHIME/FRB with hour angles up to 2.81 hr (42.1 deg \psedit{in longitude}) over a time span of three years. The 10 far side-lobe FRBs are a good sample for our science questions, as they are detected from the same telescope and the searches for repetition are performed using the same pipeline. In Section \ref{section: Far Side-lobe events}, we discuss the identification, localization, \psedit{and holographic flux calibration of the far side-lobe events. In Section \ref{sec: distances} we show the far side-lobe sample is statistically much closer than FRBs detected the main lobe and search for potential host galaxies and multi-wavelength and -messenger counterparts.} In Section \ref{section: Exposure and repetition}, we discuss the constraints on the repetition interval of one-off FRB events by using the 10 far side-lobe events. Finally, we summarize and discuss future possibilities in Section \ref{section: Conclusions}.

\section{Far Side-lobe FRBs}\label{section: Far Side-lobe events}
\subsection{Identification}

\begin{figure*}
  \centering
  \includegraphics[width=2\columnwidth]{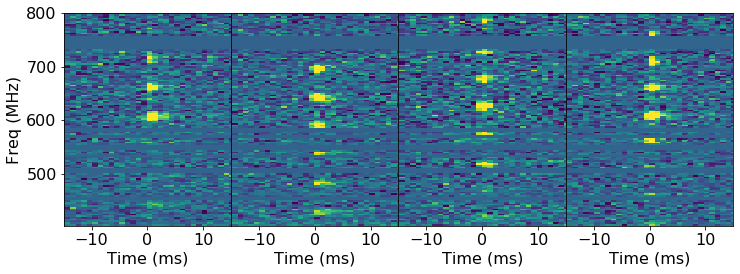}
  \caption{
    The dynamic spectrum of an example side-lobe event, FRB 20190210D. The panels from left to right represent the dedispersed dynamic spectra of each beam from East to West in the same North-South row. Each dynamic spectrum has a frequency resolution of 3.125 MHz with the range from 400 to 800 MHz and timing resolution of 0.98304 ms with a total range of 30 ms. The frequency channels dominated by RFI are masked. The colour scale is linear with intensity. 
  }
  \label{figure:1}
\end{figure*}

For CHIME/FRB, events are in far side-lobes if they are located at least several beam widths (1.3-2.6 deg from 800 to 400 MHz) away from the meridian. The signature of such a detection is a dynamic spectrum (i.e., waterfall plot, see Fig. \ref{figure:1}) with more than two spikes in the spectral profile, as has been seen in the SGR 1935+2154 detection by CHIME/FRB \citep{2020Natur.587...54C} and FRB 20110214A by the Parkes telescope \citep{2019MNRAS.482.3109P}. The detected spectrum is the product of the intrinsic spectral profile and the beam response. The beam response in the far side-lobe region shows spiky patterns across frequency channels. When a source is detected in the far side-lobe region, the spectrum of the event therefore shows at least two spikes, which is different from narrow-band events detected in the main lobe and near side-lobes \citep{2021ApJS..257...59A, 2021ApJ...923....1P}.

As CHIME/FRB forms 1024 beams with four E-W columns and 256 N-S rows \citep{2018ApJ...863...48C, 2022ApJS..261...29C}, spectral spikes arising from side-lobe events must have similar amplitude across all four East-West (E-W) beams in the same North-South (N-S) row of CHIME/FRB's formed beams. \psedit{Spikes in the spectra of FRBs could also be caused by scintillation and intrinsic emission effects. However, the spikes from the side-lobe beam response will always be of similar amplitude accross all four beams and be evenly spaced in frequency. This makes them easily distinguishable from such effects}. The sensitivity in a side-lobe is much lower than in the main lobe. Therefore, a far side-lobe event must be very bright to satisfy the triggering conditions (see Section \ref{section: Holographic calibration}.

We visually inspect dynamic spectra of CHIME/FRB events, for which the triggering criteria are described by \citet{2018ApJ...863...48C, 2021ApJS..257...59A}, and find 10 far side-lobe events from 2018 August 28 to 2021 August 31. Figure \ref{figure:1} shows the dynamic spectrum of one of the far side-lobe FRBs, while Figure \ref{fig:waterfalls_ten_sidelobe} (see Appendices) shows the dynamic spectrum of all 10 far-side lobe FRBs reported in this paper. We also show far side-lobe events from pulsars PSR B0329+54 and PSR B0531+21 (the Crab pulsar) in Figure \ref{fig:waterfalls_ten_sidelobe}, which we use to validate the following localization analysis.

\subsection{Localization}

In CHIME/FRB, there are three major localization pipelines: the header localization, the intensity localization, and the baseband localization, which provide precision on the order of degrees, sub-degree, and arcminute to sub-arcminute, respectively \citep{2018ApJ...863...48C, 2021ApJ...910..147M}.
\psedit{The header localization method in CHIME/FRB's realtime pipeline assumes that the event occurred within $\pm$ 2.5 deg of the CHIME meridian. It will therefore always provide an erroneous localization for far side-lobe events. We therefore use a different method that utilizes the CHIME/FRB intensity data. The baseband localization for the side-lobe events is under development.}


\begin{figure*}
  \centering
    \includegraphics[width=0.45\textwidth]{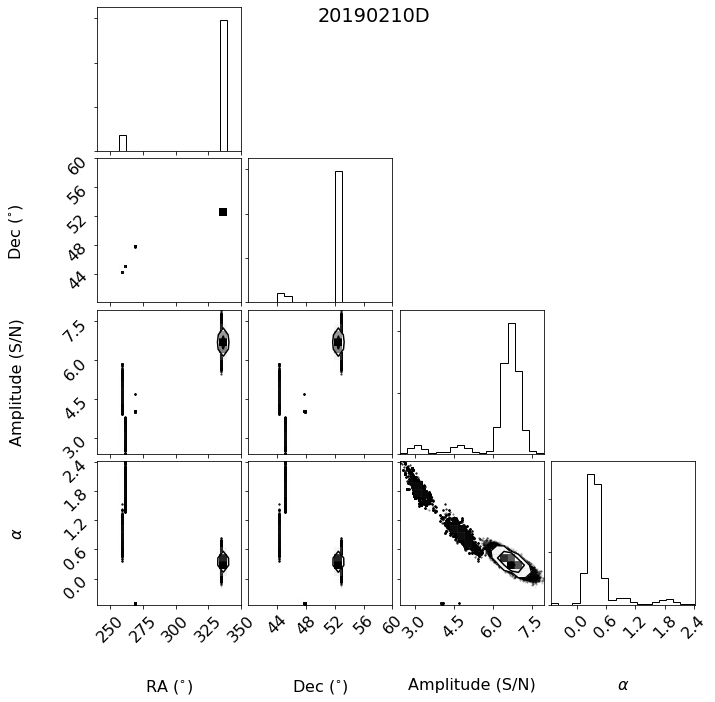} 
    \includegraphics[width=0.45\textwidth]{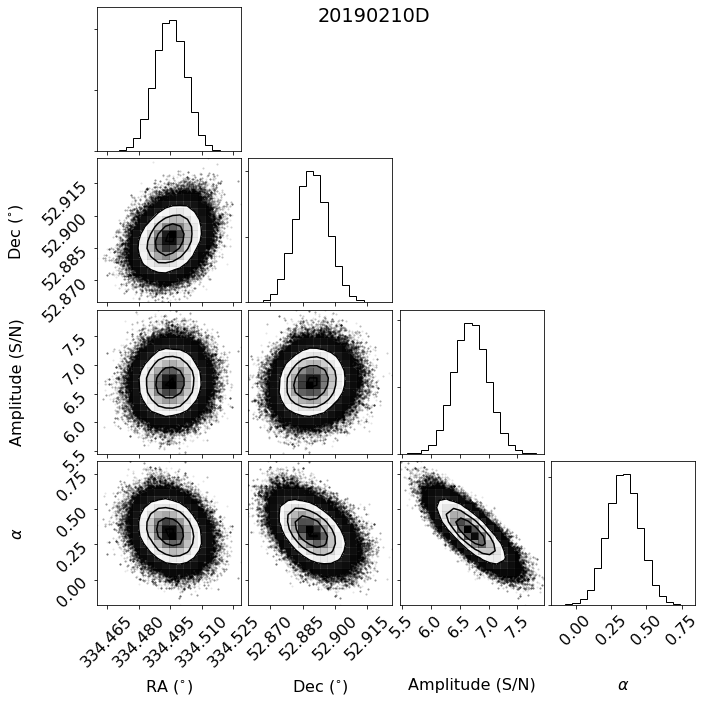} 
  \caption{MCMC output samples for the intensity localization of an example side-lobe event, FRB~20190210D. {\em Left:} Prior to filtering out sub-optimal modes in parameter space. {\em Right:} Filtered to include only the mode with the highest likelihood.
  }
  \label{figure:example_corner}
\end{figure*}

\begin{figure*}
  \centering
  \includegraphics[width=\textwidth]{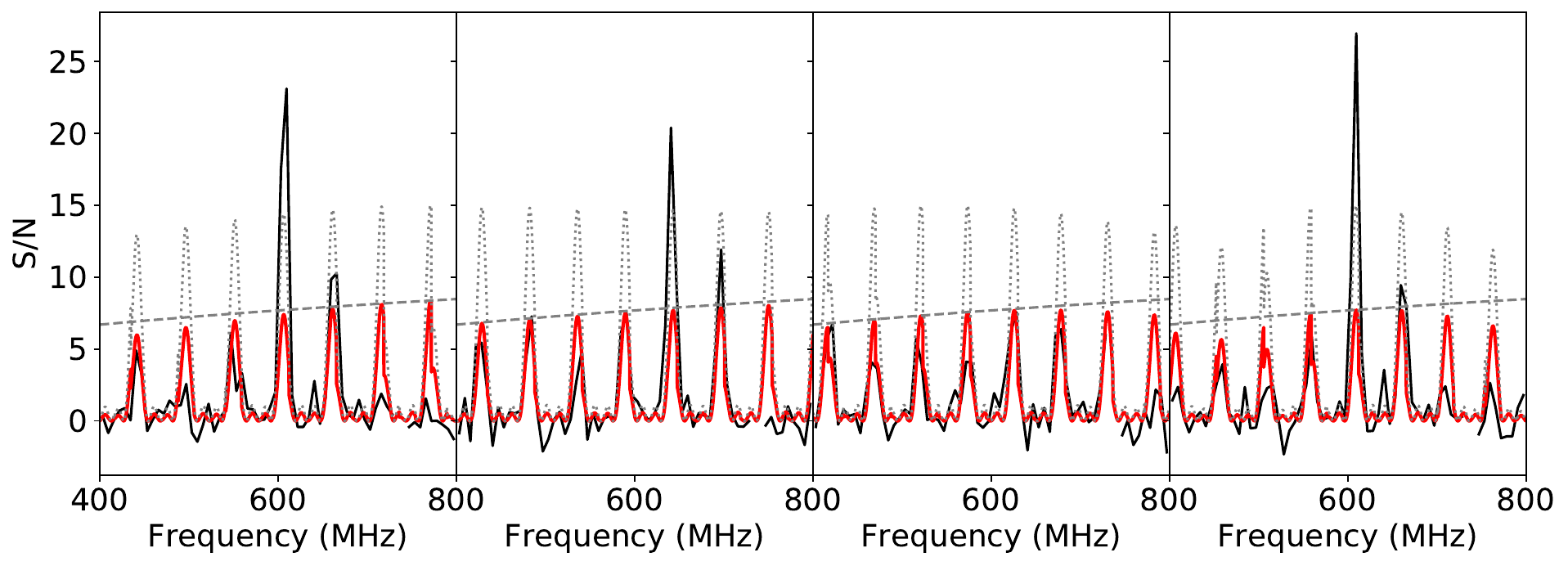}
  \caption{ Spectra (solid black) and fitted model (solid red) for the intensity localization of an example side-lobe event, FRB~20190210D. The panels represent each E-W beam as in Figure \ref{figure:1}. The model of CHIME/FRB's synthesized beam (arbitrarily scaled) is plotted as grey dotted lines and the underlying fitted power-law FRB spectrum is plotted as grey dashed lines. Note that residual spike-to-spike amplitude variations are likely due to either (i) deviations of the underlying spectrum from a power-law and/or (ii) frequency-dependent structure of the unmodeled primary beam.
  }
  \label{figure:example_spec_fit}
\end{figure*}


The CHIME/FRB intensity data consist of total-intensity dynamic spectra in 16384 frequency channels sampled at 0.98304\,ms time resolution. A dynamic spectrum is saved for each beam that detected a signal in the real-time search plus all beams adjacent (in both the E-W and N-S directions) to those detection beams \citep{2018ApJ...863...48C}. We have developed three independent methods using those data sets to localize far side-lobe events. The first method (hereafter Method 1) fits a detailed model of CHIME/FRB's synthesized beams and an underlying source spectrum to the spectra of the burst using a Markov Chain Monte Carlo (MCMC) method described below. Methods 2 and 3 attempt to simplify the localization procedure by fitting only for the spiky interference pattern in contrast to the more complex model in Method 1. These methods use the concept of diffraction and are presented in Sections \ref{subsec: method-2} and \ref{subsec: method-3} in the Appendices.

Since Method 1 shows smaller localization errors than the other two methods, we apply Method 1 for localizing far side-lobe FRBs \psedit{and describe it here}. The spectra are first downsampled to 64 subbands extracted from a time window that has four times the measured boxcar width of the burst. These spectra are fitted with the product of the CHIME/FRB beam model\footnote{Available at: \url{https://github.com/chime-frb-open-data/chime-frb-beam-model/}} \citep[described by][]{2017ursi.confE...4N,2019ApJ...879...16M} and an underlying power-law burst spectrum which can be described by
\begin{equation}\label{equation:powerlawmodel}
A_{400\,\mathrm{MHz}} \left(\frac{\nu}{400\,\mathrm{MHz}}\right)^\alpha ,
\end{equation}
where $A_{400\,\mathrm{MHz}}$ is the amplitude in units of signal-to-noise at a reference frequency of 400\,MHz, $\nu$ is the frequency, and $\alpha$ is the power-law index.
This results in four free parameters: $\alpha$,  $A_{400\,\mathrm{MHz}}$, plus the two parameters of the sky position. We use flat priors on the position of the event that span 80\,deg to either side of the meridian E–W, and in the N–S direction the extent of the beams that detected the event. Note that, for these far side-lobe events, we use only the FFT-formed beam model, and do not consider the effect of the primary beam of the telescope (which is available in the CHIME/FRB beam model\footnote{\url{https://chime-frb-open-data.github.io/beam-model/}}). This, as well as any deviation from a power-law in the true spectrum of the FRB, leads to significant residuals in the fits. This choice was made as the localization precision is dominated by the rapidly varying spike pattern of the synthesized beam, rather than the much slower varying, as a function of frequency, primary beam response. Including the primary beam model necessitates a much wider range in amplitude to be searched due to the suppression from the primary beam far from meridian.

The models are fit to the spectra using the \texttt{emcee} package \citep{2013PASP..125..306F}.\footnote{\url{https://emcee.readthedocs.io}} The output MCMC samples often display multiple modes in RA, Dec parameter space. 
However, for each of the 10 events in this paper, \psedit{the mode that has the highest posterior density is the only one for which the model visually matches the spiky signature in the data. We therefore filter out these extraneous samples as being local optima in parameter spaces that are poor fits of the data.}
Figure \ref{figure:example_corner} shows the output, before and after filtering the samples, of such a fit for an example side-lobe event. Figure \ref{figure:example_spec_fit} shows the spectra and the resulting fitted (derived from the posterior medians) model. Fitted models and MCMC output sample distributions for all of the side-lobe events are shown in Appendix Figures \ref{fig:spec_fits_ten_sidelobe}, \ref{fig:corner_plots_ten_sidelobe1}, \ref{fig:corner_plots_ten_sidelobe2}, and \ref{fig:corner_plots_ten_sidelobe3}. 

\begin{figure}
  \centering
  \includegraphics[width=0.45\textwidth]{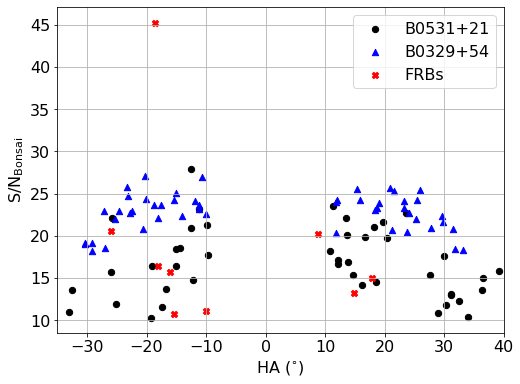}
  \caption{The S/N v.s. Hour Angle for the side-lobe events of PSR B0531+21, PSR B0329+54, and 10 FRB events.
  }
  \label{figure:sidelobes_sn_ha}
\end{figure}

\begin{figure*}
  \centering
  \includegraphics[width=0.45\textwidth]{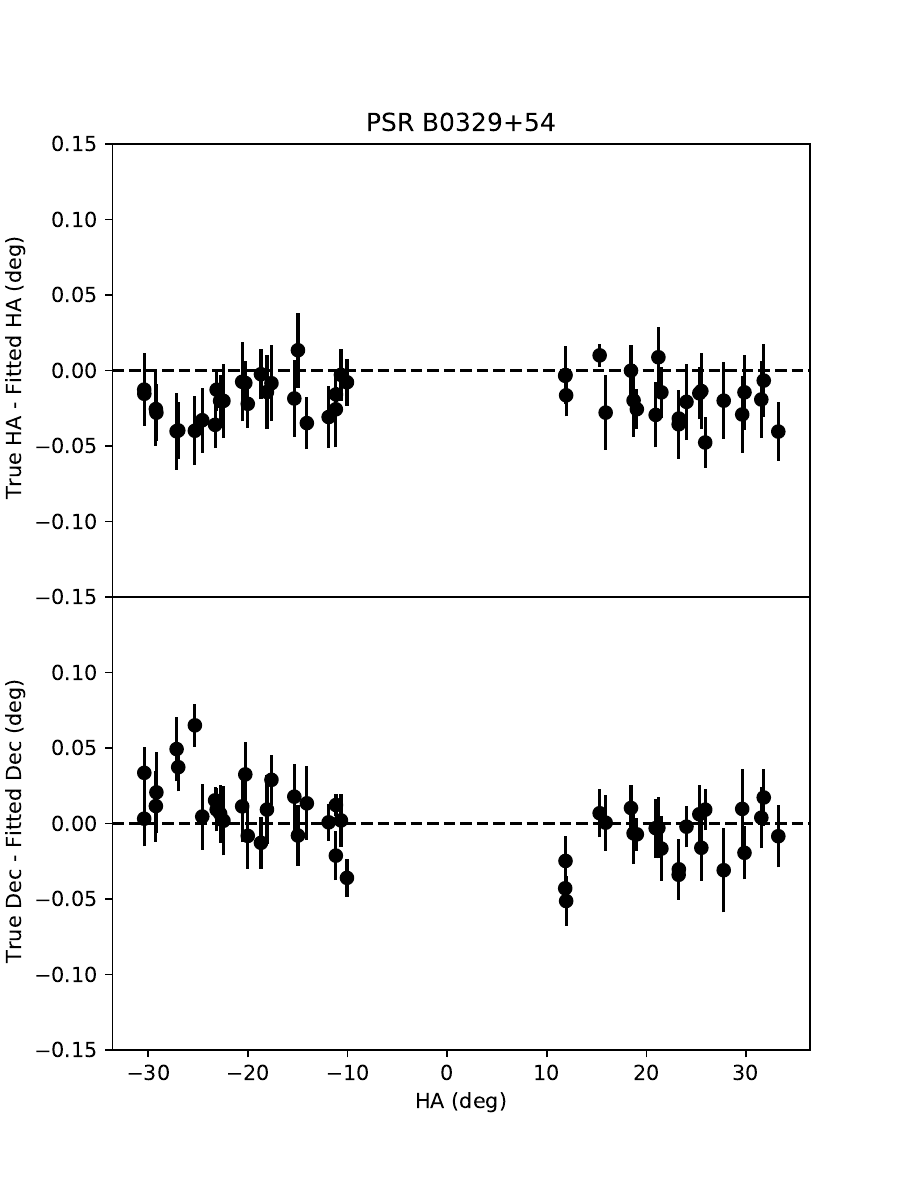}
  \includegraphics[width=0.45\textwidth]{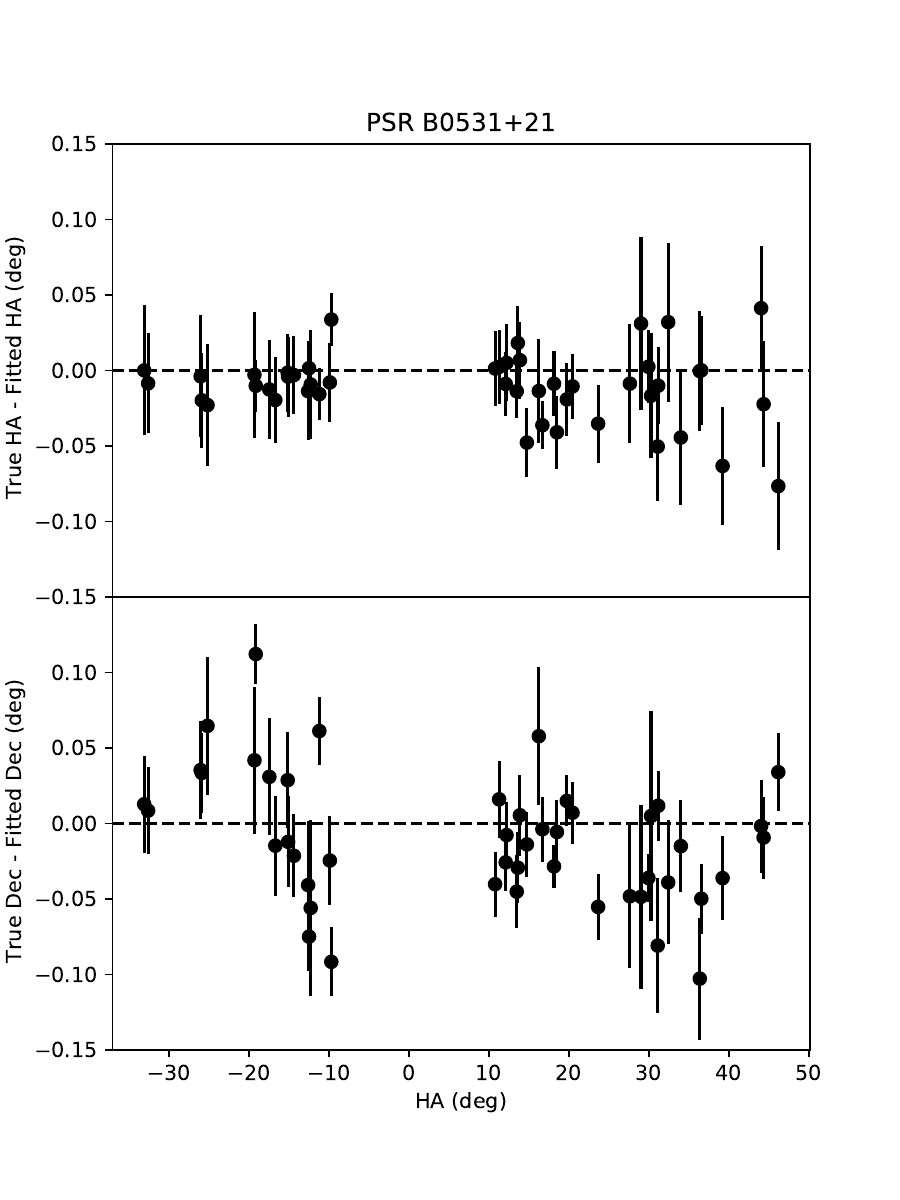}
  \caption{The localization offsets of the far side-lobe events from PSR B0329+54 and B0531+21 in terms of the range of hour angles using Method 1.}
  \label{figure:Pulsar_localization_offsets}
\end{figure*}


The giant pulses of bright pulsars PSRs B0329+54 and B0531+21 are commonly detected in the far side-lobes of CHIME and trigger the CHIME/FRB search pipeline and we use these to verify our far side-lobe localization methods. We collected intensity data of far side-lobe events from those two sources: 47 single-pulses from PSR B0531+21 and 50 single-pulses from PSR B0329+54 \citep{2005AJ....129.1993M}, with hour-angle (HA) larger than 15 deg and S/N larger than 9.0. Figure \ref{figure:sidelobes_sn_ha} shows the similar distribution of S/N and HA for the far side-lobe events from PSRs B0329+54 and B0531+21. The pulsar events therefore broadly span the E-W beam response of CHIME/FRB, so can be used to calibrate our localization methods; which primarily depend on that E-W response.

\psedit{Similar to the 10 FRB events, the posterior distributions for the test pulsars are multi-modal. For all test events the true location of the pulsar is in the mode with the highest posterior probability density and visually the additional modes are poor fits to the spiky fringe pattern. The spiky pattern is the spectral feature that is most dependent on sky location, so if the model does not match that pattern it can be safely excluded. The pulsar test events give us confidence that our choice to retain only the maximal mode in the posterior for the FRBs is robust. The outlier islands could potentially be avoided with improved priors or more sophisticated samplers that avoid local optima. However, given the evidence from the pulsar test set that the removal of these outliers is robust, we choose not to pursue it further at this time.}

Figure \ref{figure:Pulsar_localization_offsets} shows the offsets between the true position of PSRs B0329+54 and B0531+21 and the measured position from Method 1. We use these offsets to estimate a conservative systematic error for our Method 1 localizations. Namely, we use the offsets that encompass the 90\% credible intervals for 90\% of the pulsar events. This results in a systematic uncertainty in RA of $0.07\deg$ and $0.10\deg$ in Dec. In Table \ref{table: properties_table} we present this error summed in quadrature with the statistical uncertainty derived from the posterior samples. We note that there is a slight systematic offset evident in HA for both pulsars. We do not yet understand its origin, and do not attempt to correct for it as it is well-encompassed by our conservative systematic error estimation.

Appendix Figures \ref{fig:three_comparisons} and \ref{fig:two_comparisons} show the far side-lobe localization results for PSRs B0531+21 and B0329+54 from the three localization methods.

\subsection{Holographic Calibration}\label{section: Holographic calibration}

In this Section, we use holography data from three continuum sources to \psedit{estimate the S/N of the far sidelobe events had they been detected on-axis (S/N$\mathrm{_{on-axis}}$) instead of their far side-lobe location.}

Holographic techniques have been used to measure the \psedit{shape of the primary beam of the} CHIME telescope at the Dominion Radio Astrophysical Observatory \citep[DRAO][]{2016SPIE.9906E..0DB, 2022ApJS..261...29C, 2022ApJ...932..100A, 2022SPIE12190E..2VR}. Specifically, we track a bright source with the DRAO John A. Galt Telescope for at least four hours around its transit of CHIME \citep{2022SPIE12190E..2VR}. A holographic visibility is obtained as the cross-correlation of the measured voltage of a CHIME feed with that of the Galt Telescope; as the equatorially mounted Galt Telescope's response is constant, the CHIME-26m visibility provides a measurement of the CHIME beam shape, for each feed \citep{2016SPIE.9906E..0DB, 2022ApJS..261...29C, 2022ApJ...932..100A}. Before they can be used for beam analysis, the holography data undergo several processing steps.

First, the raw visibilities are fringestopped to the location of the calibrator source, removing the interferometric phase associated with the geometric delay between the point source signals arriving at CHIME and the Galt Telescope. This step isolates the complex beam phase and allows for the data to be downsampled in hour angle without decohering due to the rapid fringing in the raw visibilities.  

\psedit{The Galt Telescope is connected to the CHIME correlator by a combination of 100-m RG-214 and 305-m LMR-400 coaxial cables. CHIME, however, contains only 55\,m of coaxial cable. This disparity causes the signals from the Galt telescope to arrive at the correlator with a relative (to CHIME) delay that is a significant fraction of the integration time of a single correlation frame (2.56 $\mu$s), suppressing the amplitude of the resulting visibilities. Moreover, because the geometric delay between CHIME and the Galt telescope increases as the source transits overhead from east to west, the amount of signal loss varies (monotonically) with time, causing an asymmetry in the inferred beam response. As the decorrelation is fundamentally an effect of the finite time window used to channelize the data, we account for this by simulating the response of the polyphase filter bank implemented in CHIME. This allows us to calculate the amplitude of the decorrelation for arbitrary delays, and thus correct for it in the holography. The method of correcting for the delay is described in \citet{2022SPIE12190E..2VR}. They model two components to the delay: the ordinary geometric delay which varies with time and is completely determined by knowledge of the baseline distances and the locations of the calibrator sources, and the static delay associated with the cable length between the Galt telescope and the CHIME correlator. \citet{2022SPIE12190E..2VR} find a best-fit model which accounts for the total delay to within 0.1\%. 
We correct the holography data used in this work with this best-fit model.}

Finally, the holography data are regridded, using an inverse Lanczos interpolation scheme, onto a pre-specified grid in hour angle spanning, in the case of this analysis, from $-$60 to +60 deg, at a resolution of about 0.1 deg.

There is a potential polarization leakage issue at lower declinations and at frequency ranges above 750 MHz that is still being investigated. Additionally, the band at 725-750 MHz is dominated by radio-frequency-interference (RFI). As such we only use the holography data from three bright, high-declination calibrators, Cassiopeia A (CasA), 3C295, and Cygnus A (CygA) at 400-725 MHz for the following analysis. Their flux densities at 400$-$725 MHz are  tens (3C295) to thousands (CasA, CygA) of Jy \citep{2017ApJS..230....7P}. We list their properties in Appendix Table \ref{table: calibratos}. We normalize the raw visibility of each CHIME-26m baseline and each frequency at the meridian. To remove RFI, we cross-correlate the raw visibilities of the same calibrator in the same baseline and the same frequency channel from two different dates. We further mask frequency channels with persistent RFI and resulting in the raw squared visibility, i.e., the beam response of the CHIME feed in that holographic baseline. For each of the 32 12.5 MHz wide frequency subbands, we choose the median value of the beam response in the subband. We stack the data for the four cylinders together by choosing the median value at each subband.

\begin{figure*}
  \centering
  \includegraphics[width=2\columnwidth]{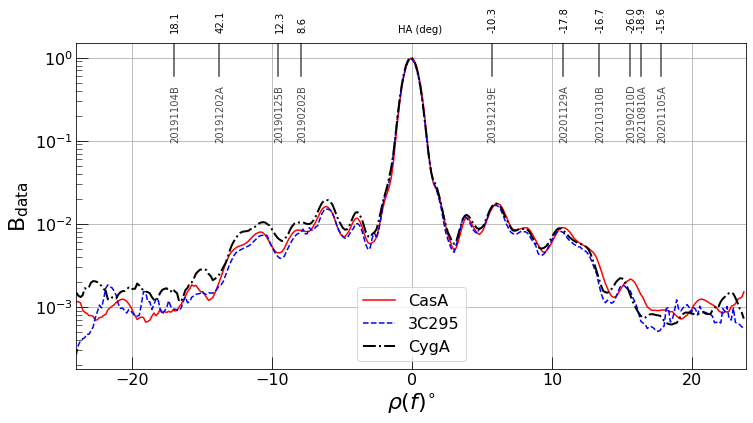}
  \caption{
   The beam response averaged over 400-725 MHz for 
   the three calibrators shown in Figure \ref{figure:beam_panels} as a function of $\rho(f)$. On top of the figure, the vertical lines mark the modeled positions of the 10 far side-lobe events in the $\rho$ coordinate that we define in Equation \ref{equation: rho_coordinate} and list in Table \ref{table: beam_reponse}, with the TNS names and the hour angle labelled. Note that 
   the peaks appear slightly offset from 0 deg, which is a virtual effect due to a resolution of 0.15 deg.
  }
  \label{figure:beam_1d}
\end{figure*}

\psedit{We then construct a model for the beam response using singular value decomposition (SVD) techniques \citep{1102314} applied to the measured cylinder-stacked beam response from the holography data. For this we use the following cylindrical coordinates}
\begin{eqnarray}
x &=& \sin(A)\cos(a),\nonumber \\
z &=& \sin(a), \nonumber \\
\rho &=& \arctan{\frac{x}{z}}, \nonumber \\
\rho(f) &=& \rho\;(\mathrm{\frac{frequency}{800 MHz}}),
\label{equation: rho_coordinate}
\end{eqnarray}
where \textit{A} and \textit{a} represent the time-dependent azimuth and altitude of the source, respectively. To get rid of the frequency-dependence of the beam size, we convert $\rho$ to $\rho(f)$ by referring to the top of the CHIME band, i.e., 800 MHz. We use the consequent beam-response in terms of frequency and $\rho(f)$ in the following analysis. Appendix Figure \ref{figure:beam_panels} shows the beam-response of the three calibrators. Figure \ref{figure:beam_1d} shows the beam-response averaged over 400-725 MHz with the position $\rho(f)$ of the 10 far side-lobe FRBs.

We use the SVD to decompose the frequency-scaled beam-response into one eigenvalue and two eigenfunctions as
\begin{eqnarray}
\mathrm{B}_{f,\rho(f)} =\sum_n {U_{fn}}{S_{n}}{V_
{n\rho(f)}^{\top}},  \\
\nonumber
\label{equation: svd}
\end{eqnarray}
where B$_{f,\rho(f)}$ is the beam response in cylindrical coordinates, with scaling the $\rho$ by a factor of frequency/(800 MHz), and for each mode $n$: $U_{fn}$ is the eigenfunction in frequency $f$, $S_{n}$ is the eigenvalue, and $V_{n\rho(f)}^{\top}$ represents the transpose eigenfunction in $\rho(f)$. Appendix Figure \ref{figure:beam_svd} shows the SVD decomposition of the beam-response of CygA. 

We use the first two modes of the B$_{f,\rho(f)}$ from CygA to reconstruct the beam model, which is shown in Figure \ref{figure:svd_model_CygA}. We compare the beam model with the B$_{f,\rho}$ from CasA, 3C295, and CygA. Appendix Figure \ref{figure:beam_svd_residuals} shows that the residuals (the absolute value of data/model) are generally around order of unity. In other words, the SVD-reconstructed beam of CygA is consistent with the holography data from the other two calibrators at high declination.

\begin{deluxetable}{lcccccc}
  \tablecaption{The measurements of beam response. \label{table: beam_reponse}}
  \tablewidth{\columnwidth}
  \tablehead{
    FRB Name\tablenotemark{a} &$\rho$\tablenotemark{b} &  \multicolumn{3}{c}{Calibrators\tablenotemark{c}} & B\tablenotemark{d}\\
      & (deg)    &  CasA & 3C295 & CygA &
    }
  \startdata
20190125B & $-$9.56	 & 0.0108 &	0.0107 & 0.0137	& 0.0117  \\
20190202B & $-$7.96	 & 0.0114 &	0.0115 & 0.0147	& 0.0125  \\
20190210D & 15.60	 & 0.0066 &	0.0061 & 0.0063	& 0.0063   \\
20191104B & $-$17.05   & 0.0048 &	0.0047 & 0.0065	& 0.0053  \\
20191201B & $-$13.77   & 0.0083 &	0.0075 & 0.0111	& 0.0090  \\
20191219E & 5.67	 & 0.0080 &	0.0087 & 0.0091	& 0.0086  \\
20201105A & 17.75    & 0.0044 &	0.0039 & 0.0043	& 0.0042  \\
20201129A & 10.81    & 0.0101 &	0.0097 & 0.0102	& 0.0100  \\
20210310B & 13.37	 & 0.0090 &	0.0083 & 0.0090	& 0.0088  \\
20210810A & 16.32	 & 0.0057 &	0.0052 & 0.0056	& 0.0055  \\
  \enddata
  \tablenotetext{a}{The TNS name of the 10 far side-lobe events.}
  \tablenotetext{b}{The best-fit position converted to cylindrical coordinates, which we define in Equation \ref{equation: rho_coordinate}}
  \tablenotetext{c}{The beam response of the three calibrators at the modeled position, for which we take average over $\rho(f)$ of the far side-lobe events.}
  \tablenotetext{d}{The averaged beam response of the three calibrators at the modeled position of the far side-lobe events.}
\end{deluxetable}

\subsection{The on-axis S/N of far side-lobe events}\label{subsection: The on-axis S/N of far side-lobe events}

We infer that the equivalent S/N of the far side-lobe events if viewed in the main-lobe is
\begin{equation}\label{equation: on_axis_SN}
\mathrm{S/N_{on-axis}} = \mathrm{\frac{G}{B}(S/N_{obs})}, 
\end{equation}
where S/N$_\mathrm{on-axis}$ is the S/N value converted to the main-lobe, G is a geometric factor in the range from 4 to 5 that \psedit{takes into account the flux lost due to the spiky synthesized beam response and is derived in Appendix \ref{app: geometric factor}}, B is the averaged \psedit{primary} beam response given in Table \ref{table: beam_reponse}, S/N$_\mathrm{obs}$ is the S/N reported by \texttt{bonsai} \citep{2018ApJ...863...48C}. 

We combine the S/N{$_\mathrm{obs}$} values from Table \ref{table: properties_table}, the beam response B from Table \ref{table: beam_reponse}, the geometric factor G of 4-5 to calculate the S/N$_\mathrm{on-axis}$ values in Table \ref{table: properties_table}. 

The median S/N{$_\mathrm{on-axis}$} of the far side-lobe events is 7700, compared to a median S/N{$_\mathrm{obs}$} for main-lobe events of 15.0 \citep{2021ApJS..257...59A}. Thus, the far side-lobe events seen by CHIME/FRB are $\sim500$ times brighter than the main-lobe events. 

\subsection{Properties of 10 far side-lobe FRBs}

We list the properties of the 10 far side-lobe FRBs in Table \ref{table: properties_table}, including the position as measured by Method 1, the observed S/N reported by \texttt{bonsai}, and the S/N had the burst been detected on-axis. For the DM, we report the total value as well as the excess DM (DM$_\mathrm{exc}$) by subtracting the NE2001 and YMW16 models at the modeled position \citep{2002astro.ph..7156C, 2017ApJ...835...29Y} from the total DM.

\begin{rotatetable*}
\begin{deluxetable*}{lcccccccccccccc}
  \tablecaption{The properties of our sample of 10 far side-lobe FRBs.}\label{table: properties_table}
  \tablewidth{\textwidth}
  \tablehead{
    FRB Name & MJD$_{400}$\tablenotemark{a}& \multicolumn{2}{c}{S/N} & {RA} & {$\sigma_{\mathrm{RA}}$\tablenotemark{b}} & {Dec} & {$\sigma_{\mathrm{Dec}}$\tablenotemark{c}} & \multicolumn{2}{c}{HA\tablenotemark{d}} & Exposure\tablenotemark{e} &DM\tablenotemark{f}             &\multicolumn{2}{c}{DM$_\mathrm{exc}$\tablenotemark{g}}   \\
     &   MJD    &    $\mathrm{obs}$    &   $\mathrm{on-axis}$   &(HH:MM:SS) & ($^\prime$) &(DD:MM:SS) & ($^\prime$) & (deg) & (hrs) & (hrs/day) &(pc\;cm$^{-3}$) & \multicolumn{2}{c}{(pc\;cm$^{-3}$)}       \\
     &      &    &          & J2000     &             & J2000     &             &       &       &           &total           & NE2001 &YMW16 
    }
  \startdata
20190125B & 58508.6275449(7) & 13.2 &5000$^{+600}_{-500}$ & 14:29:48 & 7 & +49:37:54 & 6  & 12.3 & 0.82 & 2.5 & 177.9(1) & 147.0 & 154.9  \\ 
20190202B & 58516.2854860(3) & 20.2 &7200$^{+900}_{-800}$ & 07:02:19 & 5 & +31:57:30 & 7  & 8.6 & 0.57 & 1.4 & 464.8(1) & 370.6 & 338.8 \\
20190210D & 58524.8019839(3) & 20.6 &15000$^{+1700}_{-1500}$ & 22:17:58 & 7 & +52:53:20 & 6  & -26.0&$-$1.73 & 5.7 & 359.3(1) & 106.3 & 84.6  \\
20191104B & 58791.3213754(3) & 14.9 &13000$^{+1500}_{-1300}$ & 01:20:37 & 5 & +26:42:05 & 7  & 18.1&1.20 & 2.7 & 192.2(1) & 147.1 & 156.0 \\
20191202A & 58819.0838904(3) & 14.1 &7000$^{+800}_{-700}$ & 19:51:56 & 19 &  +70:49:07 & 8 & 42.1&2.81 & 17.1 & 117.9(1) & 50.7  & 49.7  \\
20191219E & 58836.9494414(3) & 11.1 &5800$^{+700}_{-600}$ & 21:18:30 & 8 & +55:50:48 & 7  & -10.3&$-$0.69 & 2.5 &  736.7(1) & 503.0 & 336.2 \\
20201105A & 59158.2799433(1) & 10.7 &11000$^{+1300}_{-1200}$ & 02:42:18 & 5 & +14:23:13 & 7  & -15.6&$-$1.04 & 2.1 & 262.4(1) & 218.9 & 226.0 \\
20201129A & 59182.4229631(1) & 16.4 &7300$^{+900}_{-800}$ & 07:52:17 & 7 & +53:16:55 & 6  & -17.8&$-$1.19 & 4.0 & 274.6(1) & 219.6  & 221.8 \\
20210310B & 59283.3925499(1) & 15.7 &8000$^{+900}_{-800}$ & 13:42:21 & 5 & +35:33:46 & 6  & -16.7&$-$1.12 & 2.7 & 135.5(2) & 110.6 & 115.1 \\
20210810A & 59436.0351292(1) & 45.2 &37000$^{+4000}_{-4000}$ & 15:17:55 & 5 & +32:09:24 & 6  & -18.9&$-$1.26 & 3.0 & 246.9(1) & 223.4 & 223.3 \\
  \enddata
  \tablenotetext{a}{The topocentric time-of-arrival at \psedit{the CHIME site referenced to} the bottom of the band (i.e. 400.390625 MHz).}
  \tablenotetext{b,c}{\;\;\;\;The 90\%-confidence uncertainty from the Method 1 localization algorithm including systematic uncertainty (see text) in units of minutes of arc on the sky. }
  \tablenotetext{d}{The hour angle of the side-lobe event, which is relative to the meridian.}
  \tablenotetext{e}{The daily exposure time of the side-lobe events.}
  \tablenotetext{f}{The total DM of the side-lobe events, which is reported by offline algorithms via maximization of the S/N of the burst \citep{2021ApJS..257...59A}. }
  \tablenotetext{g}{The last two columns show the excess DM after subtracting the NE2001 or YMW16 model \citep{2002astro.ph..7156C,2017ApJ...835...29Y}, respectively.}
\end{deluxetable*}
\end{rotatetable*}

\section{Distances, Hosts, and Counterparts\label{sec: distances}}

\subsection{The distance of the far side-lobe events}\label{subsection: The implication for the distance of the far side-lobe events}

\psedit{The low beam response at the location of the sidelobe FRBs implies that they are much brighter than those detected in the main lobe. Here we show that this implies that they are also nearby.} As shown in the Appendix of \citet{2019Li}, the average distance of all the sources detected by instrument A with sensitivity $S_m$ will be: 
\begin{equation}
\begin{split}
		\langle r\rangle &=\frac{\int r N(r)\,dr}{\int N(r)\,dr} \\	&=\frac{\int_0^{\sqrt{E_\mr{max}/4\pi S_m}}  r^3 \pr \int_{4\pi r^2 S_m}^{E_\mr{max} } f(E)\,dE dr}
{\int_0^{\sqrt{E_\mr{max}/4\pi S_m}} r^2 \pr \int_{4\pi r^2 S_m}^{E_\mr{max} } f(E)\,dE dr}
\label{eq:rA}
\end{split}
\end{equation}
where $N(r)$ is the number of bursts visible to A at a distance $r$; $f(E)$ is the energy distribution of bursts detected from a single source; $\pr$ includes all the redshift-related evolution and selection effects, which we will discuss later; and $E_\mr{max}$ is the maximum energy of the bursts. 

Assume detector B is K times less sensitive than detector A, and therefore can detect a minimum flux of $S_m^\prime=S_m K$. 
Then we can write the average distance of the sources detected by B $\langle r^\prime\rangle$ following Equation~\ref{eq:rA} but substituting variable $r$ with $r^{\prime}=r\sqrt{K}$:
\begin{equation}
\begin{split}
\langle r^\prime\rangle
=&\frac{1}{\sqrt{K}}\times \\
&\frac{\int_0^{\sqrt{E_\mr{max}/4\pi S_m}}  r^{\prime3}\prp \int_{4\pi r^{\prime2} S_m}^{E_\mr{max}}  f(E)\,dE\,dr^\prime}
{\int_0^{\sqrt{E_\mr{max}/4\pi S_m}}  r^{\prime2}\prp \int_{4\pi r^{\prime2} S_m}^{E_\mr{max}}  f(E)\,dE\,dr^\prime} 
\end{split}
\end{equation}

For a nearby Euclidean universe, with little change of distance-related selection effects for the two instruments A \& B, $\pr=1$, or in the case of $\pr\propto r^n$, this term will cancel
in the numerator and denominator and hence 
$\langle r^\prime\rangle=\langle r\rangle/\sqrt{K}$.
The ratio of the average distance of the sources detected by detector A and B is proportional to the square root of the relative sensitivities of the two instruments and is independent of the detailed form of the luminosity function. Therefore

\begin{equation}
	\frac{\langle r^\prime\rangle }{\langle r\rangle }=
	\sqrt{\frac{S_m}{S_m^\prime}}
 \label{eq:distance-sensitivity}
\end{equation}

Since CHIME side-lobes are $\sim$500 times less sensitive than the main lobe they will therefore detect bursts on average $\sim$20 times closer than those detected in the main lobe. 

\subsubsection{The influence of redshift evolution}\label{sec:redshift evolution} 
Redshift and redshift-related evolution \psedit{and selection} effects can influence our estimates of the distance of FRBs detected with different sensitivity thresholds (Equation~\ref{eq:distance-sensitivity}). Here we show that the currently known factors have little influence on our distance and DM estimates. 

In Equation~\ref{eq:rA}, we use $\pr$ to describe the effects of redshift and redshift-related evolution. One common way to parameterize $\pr$ is $\pr=(1+z(r))^n$, where $z$ is the redshift. For the nearby universe, we can approximate $z=rH/c$, where $H$ is the Hubble constant and $c$ is the speed of light. For $z\ll1$, $\pr\approx1$ and for $z\gg1$, $\pr\approx z^n$. 
For the majority of CHIME-detected FRBs, $z$ should be between 0 and 1 \citep{2021ApJ...922...42R, 2023ApJ...944..105S}, and the side-lobe detections, $z\ll1$. Since $\prk$ is less than $\pr$ for $n>0$, $\averp<\aver/\sqrt{K}$. Therefore, the effect of redshift evolution is to make the side-lobe events closer than our estimation. When $n<0$, on the other hand, the average distance of the side-lobe events will be farther away than our estimates from the sensitivity ratio. 

There are several factors that can influence the value of $n$. The redshifts of the bursts will lead to a term $\Phi_z (r)=(1+z)^{-2-\alpha}$, where $\alpha$ is the spectral index in $F\propto \nu^\alpha$. In \citet{2023ApJ...944..105S}, the best-fit $\alpha$ from the CHIME main lobe detection is $-1.4^{+0.9}_{-1.2}$, and in this case this term will have small redshift dependence and only influence our estimate a little. 
Additionally, if the burst rate follows the star-formation rate, there will be an additional term $\Phi_\mr{SF}\approx (1+z)^{2.7}$ for $z<1$. In this case, as discussed above, the side-lobe events would be closer than our estimates. There are other redshift-related uncertainties that can be included in $\pr$. For example, when there are multiple populations of FRBs with different redshift dependencies, the $E_\mr{max}$ can also vary with respect to redshift. We will defer the discussion until we have more clues about this scenario.  

Selection effects can also influence the \psedit{distance comparison}. For example, as shown in \citet{2021ApJS..257...59A}, CHIME/FRB will miss twice as many events having DM $\simeq$ 100 pc cm$^{-3}$ compared with DM $\simeq$ 500 pc cm$^{-3}$. This can lead to the observed far side-lobe events biased towards larger DM, and hence larger distance. However, the change of DM selection function is smooth near the observed DM range of side-lobe events, therefore its influence on $\averp$ \psedit{is minor}. 

On the other hand, the selection effects on the burst width are large, with bursts longer than 10 ms rarely detected. For specific scenarios, such as FRBs scattered by foreground galactic halos, the scattering time of bursts can be largely related to their distances, resulting in a large fraction of bursts with $z>1$ undetected. The estimate of $\aver$ from the observed events will then be much lower than the actual value, and this can result in a substantial underestimate of the DM of the side-lobe events. However, in this halo scattering scenario, there will be a strong correlation between the observed DM and the scattering time, which is not seen in current FRB samples \citep{2022ApJ...927...35C}. 

\psedit{As determined in Section \ref{sec: distances}, the far side-lobe events should be $\sim$20 times closer than the main-lobe events as a population. Though the aforementioned redshift-dependant effects may change this distance ratio slightly, it is clear that the far sidelobe events should be preferentially nearby.}

\subsection{Host galaxy search}
\label{sec:host}

\psedit{Since the far sidelobe FRBs are statistically closer that those detected in the main lobe they should be especially interesting for host galaxy studies. We} therefore search for host galaxies of the 10 side-lobe events within the reported 90\% confidence localization region listed in Table \ref{table: properties_table}. Prior to that, we check if there are cataloged Galactic \ion{H}{2} and/or star-forming regions \citep{anderson2014wise, avedisova2002catalog} within the localization region of the side-lobe events that can contribute to their extragalactic DMs listed in Table \ref{table: properties_table}. Only in the case of FRBs 20190210D and 20191219E, we identify multiple nearby ionizing regions. This is unsurprising as both the side-lobe events are Galactic plane sources with Galactic latitudes (b) of $-$3.3 and 4.5 degrees, respectively. The contribution of these ionizing regions to the FRB DM is hard to quantify due to the poor localization region of FRBs 20190210D and 20191219E. This, along with considerable uncertainty in the predictions of Galactic DM models for low latitude sources \citep[b $\lesssim$ 10 degrees;][]{2021PASA...38...38P}, make the host association quite challenging \citep[for example, see][]{2019ApJ...885L..24C} unless the FRBs are localized to arcsecond precision \citep{2023arXiv230101000R}.  

Due to the large localization region ($\approx$ 36 arcmin$^{2}$), we are unable to make a robust host association for any of our side-lobe FRBs. Although the standard formalism of chance association probability (P$_{\rm cc}$) described in \cite{2017ApJ...849..162E} would give $\lesssim$ 10\% probability only for galaxies of r-band apparent magnitude m$_{\rm r} \leq$ 13 AB mag in the localization region of side-lobe events, this is not a sufficient condition to make a robust association \citep[for example, see][]{2021ApJ...919L..24B}. Noting the prospects of some of the side-lobe events to be local Universe FRBs \citep{2022PASP..134i4106L}, we check the Galaxy List for the Advanced Detector Era Version 2.3 (GLADE v2.3) catalog \citep{2018MNRAS.479.2374D}, which contains all of the brightest galaxies up to a luminosity distance of 91 Mpc, to identify galaxies that satisfy the aforementioned r-band constraint. In all except FRBs 20190125B and 20210310B, we do not find a very nearby ($<$ 100 Mpc) galaxy within the FRB 90\% localization region. In the FRB 20190125B 90\% confidence localization region, we find NGC 5660, a star-forming spiral galaxy at a distance of 38 Mpc \citep{1983ApJS...52...89H}, as a promising host candidate (P$_{\rm cc} < 1$\%), and in the case of FRB 20210310B, we find NGC 5273 \citep[16 Mpc][P$_{\rm cc} < 1$\%]{2017ApJ...850...74K} and NGC 5276 \citep[84 Mpc][P$_{\rm cc}\sim 10$\%]{2012ApJS..203...21A} as promising candidates. \psedit{Note that with 10 FRBs in our sample, the probability of seeing an association with P$_{\rm cc}< 1$\% would be $\sim$10\%. We therefore consider these host candidates promising and do not claim a firm association.} If the FRB source is in a globular cluster, as is FRB 20200120E which has been localized to a globular cluster $\sim$15 arcmin away from the center of M81 \citep{2021ApJ...910L..18B, 2022Natur.602..585K}, such an association could be missed during the search. 

\subsubsection{FRB 20191202A}\label{sec:frb20191202a}

FRB 20191202A is the most interesting source among the 10 side-lobe events because of the very low DM$_{\rm ext} \approx$ 50 pc cm$^{-3}$ (see Table \ref{table: properties_table}). Using the technique described by \cite{2021ApJ...919L..24B}, we estimate the maximum redshift z$_{\mathrm{max}}$ of the FRB to be $\approx$ 0.04 (90\% confidence upper limit). If FRB 20191202A is located at z$_{\mathrm{max}}$, and if we assume it is in a faint star-forming dwarf galaxy similar to that of FRB 20121102A \citep[$\mathrm{M_{r}} = -17$ AB mag;][]{2017ApJ...834L...7T}, it would have an r-band magnitude of $\approx~$19.9 AB mag. As the FRB field-of-view is imaged by the Panoramic Survey Telescope and Rapid Response System (Pan-STARRS) release 1 (PS1) Survey \citep{2016arXiv161205560C} with an r-band depth for galaxies $\approx$ 21.5 AB mag (5$\sigma$), we search for extended galaxy candidates in the Pan-STARRS1 Source Types and Redshifts with Machine learning (PS1-STRM) photometric redshift (z$_{\rm ph}$) catalog \citep{2021MNRAS.500.1633B} and find 66 sources. When we apply the r-band Kron magnitude (rKmag) $<$ 19.9 AB mag and z$_{\mathrm{photz}} - 3\sigma_{\mathrm{photz-err}} <$ z$_{\mathrm{max}}$ = 0.04, we find four galaxies, which are listed in Table \ref{tab:20191202A}. If none of the four galaxies is the FRB host, it would mean that the FRB host is the faintest host known to date. 

Interestingly, we note that the FRB excess-DM is similar to that of FRB 20200120E \citep{2021ApJ...910L..18B}, where the source is localized to a globular cluster of M81 at 3.6 Mpc \citep{2022Natur.602..585K}. Therefore, provided the FRB has an extragalactic origin, the absence of a very nearby host in the GLADE 2.3V catalog within its completeness limit of $\approx$ 100 Mpc suggests that either the predicted Milky Way DM contribution is significantly overestimated or the FRB host contribution is negligible. Within $\sim$ 3-4 degrees centered at the FRB location, we find several Galactic pulsars\footnote{Pulsars J1955+6708, J1953+67, and J2043+7045, identified using Pulsar Survey Scraper \citep{2022ascl.soft10001K} (visited on 15/02/2023).} with DM $\approx$ 57 pc cm$^{-3}$. Moreover, from \cite{ocker2020ApJ}, the maximum DM through the Milky Way’s  disk at the FRB Galactic latitude of $\approx$ 21 deg. is $66 \pm 7$ pc cm$^{-3}$. 
These measurements suggest that the NE2001 and YMW16 models do not significantly overestimate the Milky Way disk DM contribution, making the FRB particularly promising for constraining the Milky Way halo DM contribution \citep[see][]{2023arXiv230103502C}. 

\begin{table}[ht]
\caption{Galaxies with rKmag $<$ 19.9 AB mag in the FRB 90\% confidence localization region with z$_{\mathrm{photz}} - 3\sigma_{\mathrm{photz-err}} < $ z$_{\rm max}$.}
\label{tab:20191202A}
\begin{center}
\begin{tabular}{@{} *6c @{}}
\toprule
Number & R.A. &Dec.& rKmag& z$_{\mathrm{photoz}}$&$\sigma_{\mathrm{photoz-err}}$\\
&J2000& J2000& AB mag.& &\\ \midrule 
1 & 298.0218 & 70.7713 & 19.32 & 0.05 & 0.03\\
2 & 298.1829 & 70.8063 & 19.60 & 0.18 & 0.06\\
3 & 297.6873 & 70.8567 & 19.11 & 0.08 & 0.04\\
4 & 297.5976 & 70.8551 & 19.65 & 0.25 & 0.1\\
\bottomrule 
 \hline
\end{tabular}
\end{center}
\end{table}

\subsection{GRB and GW Counterparts search}

Using our localizations, we check for possible high-energy counterparts to the 10 side-lobe FRBs. More specifically, we check for temporal (up to one week), and spatial (within 3$\sigma$ of each other's localization regions) coincidence between our set of FRBs and all known gamma-ray bursts (GRBs) published in the Gamma-ray Coordination Network (GCN)\footnote{\url{https://gcn.gsfc.nasa.gov}} circulars. We limit the GRBs to those that are well localized (e.g., localization errors $<$1 deg in RA and DEC), as it is difficult to claim significant spatial coincidences for GRBs with either unknown or large uncertainty regions. We do not find any GRB-FRB pairs with the given criteria. 
If we only search for spatial (rather than temporal and spatial coincidence), we similarly do not find any coincident GRB-FRB pairs \citep{2022arXiv220800803C}. 

We similarly checked the LIGO GraceDB\footnote{\url{https://gracedb.ligo.org/superevents/public/O3/}} for gravitational wave (GW) events with temporal (within one week) and spatial coincidence.  Since the FRB localization region is much smaller than the LIGO error region, we define a spatial coincidence to be when the FRB position is within the 90\% localization region of the GW event. We also restrict the search to events that involve at least one neutron star since a pure binary black hole merger is not expected to create electromagnetic bursts \citep{2018PhRvD..98l3016D}. We further restrict the false alarm rate (FAR, as mentioned in \citet{2016PhRvD_ligo_far_ref, 2017ApJ...849..118N}, a measurement of how frequently a non-astrophysical event would be falsely reported from the GW data searches), to FAR $< 10^{-7}\,\mathrm{yr^{-1}}$ and select only vetted superevents (labelled \texttt{ADVOK}). Of our detections, no events were both spatially and temporally coincident with a GW event. FRB 20191219E was temporally coincident with a GW event $–$ GW S191213g, which has a 77\% chance of being a binary NS merger and 23\% chance of being terrestrial noise. However, the two were not spatially coincident. Hence we consider that these two events are unlikely to be linked. 

\section{Repetition and Exposure}\label{section: Exposure and repetition}

In this Section, we describe the search for repeat bursts from the far sidelobe FRBs, and calculate the lower bound of the exposure time and the lower bound of the repetition interval. 

\subsection{The search for repetition}
\label{subsect:repetition_search}

Since the side-lobe FRBs were sufficiently bright to be detected in the side-lobe region, any repeating bursts from their sources above the detection threshold could potentially be detected either in the main lobe or the side-lobe. To search for repetition in the main lobe from 2018 August 28 to 2021 August 31 in the database, we apply the following conditions:  
\begin{enumerate}
  \item the S/N of the trigger must be higher than 9,
  \item the position of the trigger must be less than 3 deg (i.e., within the main beam) away from the modeled position of the side-lobe event, 
  \item the absolute difference of the DM between the trigger and the side-lobe event must be less than 5 pc\;cm$^{-3}$, as CHIME known repeaters' DM variation is less than 5 pc\;cm$^{-3}$ \citep{2019ApJ...885L..24C, 2020ApJ...891L...6F, 2023arXiv230108762T}.
\end{enumerate}

The conditions for the initial trigger for which we apply these criteria is described by \citet{2021ApJS..257...59A}. We do not find any associated event detected in the main lobe. 

\begin{figure}
  \centering
  \includegraphics[width=0.45\textwidth]{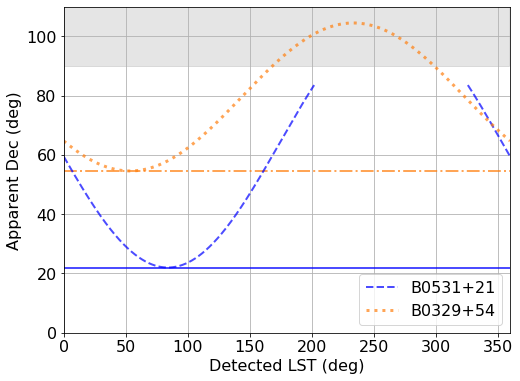}
  \caption{The apparent declinations of PSRs B0531+21 and B0329+54. The dashed and dotted lines represent the apparent positions of the two pulsars. PSR B0531+21 is below the horizon at CHIME when the apparent declination is higher than 83.5 deg. The horizontal line marks the true declination of the pulsar, when the detected LST is equal to the RA of the pulsar. The source is in a lower transit when the declination of the apparent curve is higher than 90 deg, which is marked in grey. 
  }
  \label{figure:apparent_curves}
\end{figure}

To investigate whether there are repeating events detected in the side-lobes, we use the modeled position of each side-lobe event to construct the apparent position on the CHIME sky as the apparent curve, and we searched for potentially associated events along the apparent curve. Figure \ref{figure:apparent_curves} shows the apparent curves for PSRs B0531+21 and PSRs B0329+54 as a function of the detected local sidereal time (LST) at CHIME. The transit happens when the LST is equal to the RA of the source.

We search for repetition of the 10 side-lobe FRBs in all CHIME/FRB events with the same conditions as above. If the associated event is again in the side-lobe region and on the apparent curve (i.e., the hour angle is several degrees away from the meridian), we expect a trigger would show the spiky pattern in the waterfall plot. However, we did not find candidates that satisfy these conditions. One possible selection bias is that the spectral bandwidth of the repeating event is less than the separation of the spikes in the dynamic spectrum, which we may miss during the virtual inspection. Table \ref{table: properties_table} shows that the minimum of the absolute hour-angle of the 10 far side-lobe events is 8.6 deg. The corresponding spectral separation of the spikes is $\sim$90.6 MHz (See Method 2 in Appendix), where 5 of the 62 repeater events reported in the first CHIME/FRB catalog have a spectral bandwidth less than 90.6 MHz \citep{2021ApJS..257...59A}. Since we compare the spiky separation in terms of the minimum of the hour-angle offset and the spectral bandwidth from 62 repeat events, the probability of missing the narrow-band repeating events in the side-lobe is 5$/$62 $\sim$8$\%$.

Subject to the above caveats, we conclude that the CHIME telescope probably did not detect repetitions from any of the 10 side-lobe FRBs from 2018 August 28 to 2021 August 31. 

\subsection{The lower bound of the exposure time and the repetition interval}\label{subsect:exposure_time}

In Section \ref{section: Holographic calibration} we show that the sensitivity at the hour angle of the detection is lower than the sensitivity interior to the hour angle of the detection (see Figure \ref{figure:beam_1d}). Hence, we consider twice the hour angle of each detection as a lower bound of the daily exposure time for each of the repetition intervals, i.e., we only consider the region with higher sensitivity than the sensitivity at the detecting hour-angle for the exposures. \citet{2022ApJ...932..100A} use solar data to show that CHIME’s response is highest near the meridian \citep[see Figure 3 in][]{2022ApJ...932..100A}. Regardless of whether the source is in the main lobe or side-lobe, the detection criterion is that the triggering S/N (i.e., the product of the beam-response and the S/N ratio at the center of the main-beam) is 9 and above.   

Note that the exposure time in the main lobe is defined as having the sky location in question within the full-width at half-maximum (FWHM) region of a synthesized beam at 600 MHz \citep{2021ApJS..257...59A}, which does not include locations between the FWHM of beams. However in the far side-lobes, since the beam response is spiky, for sufficiently broadband FRBs (but see Section \ref{subsect:repetition_search}), there should be sufficient sensitivity at all locations.

To constrain the repetition interval we apply Poisson statistics
\citep[following the formalism of][]{2018MNRAS.475.5109O}. For each of the far side-lobe events, 
\begin{equation}\label{equation: poisson_rate_single}
\mathrm{P_{i}(k=0;\lambda_{i})} = \mathrm{e^{-\lambda_{i}}}, \mathrm{\lambda_{i}=r_{i}t_{i}},
\end{equation}
where $\mathrm{P_{i}(k=0;\lambda_{i})}$ represents the individual Poisson probability distribution of zero repetition
: $\mathrm{k}$ is the number of occurrences in an interval $\mathrm{t_{i}}$ and we take zero for non-repeating sources, $\mathrm{\lambda_{i}}$ is the the individual average number of events and equal to the individual repetition rate ($\mathrm{r_{i}}$) times the individual observing duration ($\mathrm{t_{i}}$).\\
Since there is no repetition from each of the far side-lobe events, if we assume that each of them has the same lower bound of the repetition interval (i.e., $\mathrm{1/r = 1/r_{1} = ... = 1/r_{10}}$), the Poisson distribution for all far side-lobe event is 
\begin{equation}\label{equation: poisson_rate_total}
\mathrm{P_{tot}(k=0;\lambda_{tot})} = \mathrm{\prod_{i=1}^{10} e^{-r_{i}t_{i}} = e^{-r(\sum_{i=1}^{10}{t_{i}})}=e^{-rT_{exp, total}}},
\end{equation}
where $\mathrm{T_{exp, total}}$ is the total exposure time. Hence, we can sum individual exposure times into a total exposure time for the far side-lobe samples.
    
Since CHIME only observes a strip of sky transiting directly overhead, the exposure varies significantly with declination. We account for this dependence by scaling the exposure with the cosine of the declination \citep{2022PhDT........19C}. Hence, the daily exposure time for each side-lobe event, as shown in Table \ref{table: properties_table}, is

\begin{equation}
\mathrm{T_{\mathrm{exp, daily}}} = 2\times|\mathrm{HA}|\;\mathrm{(deg)}
\times\frac{1}{\mathrm{\cos{\theta_{DEC}}}}\times\mathrm{\frac{24\;hours}{360\;deg}}\;,\label{equation: daily_exposure_time}
\end{equation}
where the factor of 2 accounts for the two sides of side-lobe, $|\mathrm{HA}|$ and $\mathrm{\theta_{DEC}}$ corresponds to the absolute HA and the modeled declination value of each far side-lobe event in Table \ref{table: properties_table}.

From 2018 August 28 to 2021 August 31, the CHIME/FRB system was operational for 845.59 out of 1099 days, and on average 988.6 of 1024 online synthesized beams were running during this up time. This leads to 74$\%$ operational up time\footnote{The relevant beams were operational the average amount of time.}. Thus, we calculate the lower bound of the total exposure time for the 10 side-lobe event as
\begin{equation}
\begin{split}
\mathrm{T_{\mathrm{exp, total}}} &= 0.74\times\mathrm{\sum^{10}_{i=1}}\;\mathrm{T_{\mathrm{exp, daily}}}
\times1099\;\mathrm{days}\\
&\simeq 0.74\times43.75\;\mathrm{\frac{hours}{day}}
\times1099\;\mathrm{days}\\
&={35580\;\mathrm{hrs}}.\label{equation: exposure_time}
\end{split}
\end{equation}
With a total exposure time of 35580 hours\footnote{Note that the exposure time is dominated by FRB 20191202A, which has a high declination of $\sim$71 deg and a daily exposure time of 17.1 hours, as listed in Table \ref{table: properties_table}.} for the 10 side-lobe events, CHIME/FRB did not detect repeat bursts from the 10 far side-lobe events listed in Table \ref{table: properties_table}. 

Note that by using twice the hour-angle for our exposures we are being quite conservative. The sensitivity of CHIME's primary beam far from meridian is fairly flat (see Figure \ref{figure:beam_1d}), i.e., it does not drop rapidly as a function of hour angle. There is therefore significant sensitivity outside of our exposure window that is comparable to that within the window. There may also be unaccounted-for sources of incompleteness, such as our search for repetition missing narrowband bursts (Section \ref{subsect:repetition_search}).

Since the exposure time is dominated by the low-sensitivity side-lobes with $\sim$ 10$\%$ from the high-sensitivity main lobe, the non-detection of repeat bursts implies that follow-up observations of CHIME/FRB non-repeating sources with a high sensitivity telescope have a significant chance of non-detections. For instance, the sensitivity ratio\footnote{The side-lobe sensitivity is $\sim$1$\times10^{-2}$--1$\times10^{-3}$ lower than the main lobe (Figure \ref{figure:beam_1d}). Since there are four E-W beams, we consider the side-lobe sensitivity is at least $\sim$4$\times10^{-2}$ less than the main lobe.} between CHIME/FRB's side-lobe region and main lobe is $\sim$4$\times10^{-2}$, which is approximately equal to the sensitivity ratio between CHIME/FRB's main lobe and FAST's main beam. For the total exposure time of 35580 hours on the 10 far side-lobe FRBs, CHIME/FRB has $\sim$32000 hours and $\sim$3580 hours exposure time in the side-lobe and main-lobe, respectively.  To have the same monitoring efforts on \psedit{CHIME/FRB main-lobe non-repeaters}, FAST would need $\sim$400 hours exposure time, which would be prohibitively expensive to perform on the large sample of non-repeating FRBs. On the other hand, the future detection of even one repeat burst of our 10 sidelobe events by any telescope, including FAST, would be interesting, and would strongly suggest universal repetition with a wide range of repeat times. \psedit{Follow-up observations of the smaller sample of 10 side-lobe events could therefore be more fruitful than the much larger sample of main-lobe non-repeaters.}. 

Using Equation \ref{equation: poisson_rate_total} with a confidence level (CL) of 95$\%$ and a total exposure time of 35580 hrs, the corresponding lower bound on the repetition interval (1/$\mathrm{r}$) above CHIME/FRB's sensitivity limit is 11880 hours. This limit could explain the non-detection of repeating bursts with tens to hundreds of hours follow-up observations by the Parkes \citep{2019MNRAS.482.3109P, 2015ApJ...799L...5R, 2016Sci...354.1249R}, GBT \citep{2015Natur.528..523M}, and Arecibo \citep{2022arXiv220409090G} telescopes. 

\psedit{Applying Equation \ref{equation: poisson_rate_total} to the CHIME/FRB main-lobe non-repeater sample, which has a total exposure time of 25700 hours \citep{2021ApJS..257...59A}, results in a lower bound on the repetition interval of 8580. This is similar, though not as constraining as the bound placed using our far-sidelobe sample. The two bounds probe in very different regimes: a large number of sources with short exposure times and a small number of bright sources with extremely long exposure times.}

Whether all FRBs repeat or not is an open question in the FRB field. In the first CHIME/FRB catalog, the repeaters and the apparent one-off events show different properties in morphology, where in general the former are narrow-band and the latter are broadband \citep{2021ApJS..257...59A, 2021ApJ...923....1P}. Here we compare the repetition intervals of repeaters to that of the far side-lobe events. The average repetition rate of 44 CHIME/FRB repeaters is $\sim$0.2 hr$^{-1}$, which leads to a mean repetition interval of 5 hours\footnote{Note the median repetition interval is $\sim$ 33 hours.} \citep{2021ApJS..257...59A, 2023arXiv230108762T}. 

Our measured mean repetition interval for the 10 side-lobe events, 11880 hours, is 2380 times longer than that for the CHIME/FRB repeater sample. Under our assumptions, this implies that these two samples come from vastly different regimes in repetition rate. These differing regimes may be due to a wide, or multi-modal, distribution of rates in a single physical population, or it may be due to separate physical populations, one of which may be cataclysmic. Future observations with longer exposure time on FRBs, such as CHIME/FRB \citep{2018ApJ...863...48C} and BURSTT \citep{2022PASP..134i4106L}, would be helpful to further constrain the repetition of the apparent non-repeating FRBs.

\psedit{Note that the repetition distribution of many known repeaters are not Poissonian. The deviation from Poisson statistics seems to take two forms: clustering on short timescales \citep[$<$1\,s;][]{2021ApJ...920L..23Z,2021ApJ...922..115A} and the emission of bursts in active windows \citep{2023ApJ...947...83C}, some of which are periodic \citep[FRBs 20121102A and 20180916B;][]{2020Natur.582..351C,2020MNRAS.495.3551R}. On long timescales and within their active windows, repeating FRBs appear to be consistent with Poisson repetition \citep{2021MNRAS.500..448C,2023ApJ...956...23S}. Assuming Poisson-distributed waiting times in our above analysis is therefore a valid approximation. The Poisson rates considered here are related to the rates during active windows by the duty cycle of a source's active windows. In this context, the non-detection of repetition could be due to a low duty cycle for active windows, i.e. a source could have a high burst rate while it is active but with rare periods of activity.}

\section{Conclusions}\label{section: Conclusions}

We report 10 far side-lobe FRBs detected by CHIME/FRB from 2018 August 28 to 2021 August 31. We use the intensity data for these sources to localize them with sub-degree precision. \psedit{These FRB sources detect in the far sidelobes should be drawn from a population that is about $20\times$ closer than those in the mainlobe sample. They are thus interesting targets for host galaxy and counterpart follow-up. Indeed, FRBs 20190112B and 20210310B are associated with potential host galaxy candidates at 38 and 16 Mpc, respectively, with P$_{\rm cc} < 1$\%}. Over three years, we did not find any repeat bursts from any of the side-lobe sources from the CHIME/FRB database with conditions of S/N $\geq$ 9.0, distance between the modeled position of the far side-lobe event and the header position of the CHIME/FRB database less than 3 deg, and DM difference larger than 5 pc\;cm$^{-3}$.

With the long exposure time of 35580 hours on far side-lobe events, we find that the Poisson repetition interval for the one-off events is longer than 11880 hours, which is at least 2380 times longer than for CHIME/FRB repeaters. Longer exposure time on FRBs with future FRB-surveys would be helpful to understand whether all FRBs repeat or not. This study shows the advantage of considering events detected in the low-sensitivity sidelobes of telescopes to probe for rare, bright events with a long exposure time. \\

We acknowledge that CHIME is located on the traditional, ancestral, and unceded territory of the syilx/Okanagan people. We thank the Dominion Radio Astrophysical Observatory, operated by the National Research Council Canada, for gracious hospitality and expertise. CHIME is funded by a grant from the Canada Foundation for Innovation (CFI) 2012 Leading Edge Fund (Project 31170) and by contributions from the provinces of British Columbia, Quebec and Ontario. The CHIME/FRB Project is funded by a grant from the CFI 2015 Innovation Fund (Project 33213) and by contributions from the provinces of British Columbia and Quebec, and by the Dunlap Institute for Astronomy and Astrophysics at the University of Toronto. Additional support was provided by the Canadian Institute for Advanced Research (CIFAR), McGill University and the McGill Space Institute via the Trottier Family Foundation, and the University of British Columbia. The Dunlap Institute is funded through an endowment established by the David Dunlap family and the University of Toronto. Research at Perimeter Institute is supported by the Government of Canada through Industry Canada and by the Province of Ontario through the Ministry of Research $\&$ Innovation. The National Radio Astronomy Observatory is a facility of the National Science Foundation (NSF) operated under cooperative agreement by Associated Universities, Inc. FRB research at UBC is supported by an NSERC Discovery Grant and by the Canadian Institute for Advanced Research. FRB research at WVU is supported by an NSF grant (2006548, 2018490). Computations were performed on the Niagara and Cedar supercomputers at the SciNet HPC Consortium \citep{2010JPhCS.256a2026L,2019arXiv190713600P}. SciNet is funded by: the Canada Foundation for Innovation; the Government of Ontario; the Ontario Research Fund - Research Excellence; and the University of Toronto.

P.S. is a Dunlap Fellow. 
Ue-Li Pen receives support from Ontario Research Fund—research Excellence Program (ORF-RE), Natural Sciences and Engineering Research Council of Canada (NSERC) [funding reference number RGPIN-2019-067, CRD 523638-18, 555585-20], Canadian Institute for Advanced Research (CIFAR), Canadian Foundation for Innovation (CFI), the National Science Foundation of China (Grants No. 11929301), Thoth Technology Inc, Alexander von Humboldt Foundation, and the Ministry of Science and Technology(MOST) of Taiwan(110-2112-M-001-071-MY3). Computations were performed on the SOSCIP Consortium’s [Blue Gene/Q, Cloud Data Analytics, Agile and/or Large Memory System] computing platform(s). SOSCIP is funded by the Federal Economic Development Agency of Southern Ontario, the Province of Ontario, IBM Canada Ltd., Ontario Centres of Excellence, Mitacs and 15 Ontario academic member institutions. 
MB is a Mcwilliams Fellow and an International Astronomical Association Gruber fellow. 
A.P.C. is a Vanier Canada Graduate Scholar.
K.R.S acknowledges support from FRQNT Doctoral Research Award.
S.P.T. is a CIFAR Azrieli Global Scholar in the Gravity and Extreme Universe Program. 
B.\,C.\,A. is supported by an FRQNT Doctoral Research Award.
A.M.C. was supported by the Government of Ontario through an Ontario Graduate Scholarship. 
M.D. is supported by a CRC Chair, NSERC Discovery Grant, CIFAR, and by the FRQNT Centre de Recherche en Astrophysique du Qu\'ebec (CRAQ).
F.A.D. is funded by the U.B.C Four Year Fellowship.
G.E. is supported by an NSERC Discovery Grant (RGPIN-2020-04554) and by a Canadian Statistical Sciences Institute (CANSSI) Collaborative Research Team Grant. 
B.M.G. is supported by an NSERC Discovery Grant (RGPIN-2022-03163), and by the Canada Research Chairs (CRC) program. 
V.M.K. holds the Lorne Trottier Chair in Astrophysics \& Cosmology, a Distinguished James McGill Professorship, and receives support from an NSERC Discovery grant (RGPIN 228738-13), and from the FRQNT CRAQ. 
C.L. was supported by the U.S. Department of Defense (DoD) through the National Defense Science \& Engineering Graduate Fellowship (NDSEG) Program. 
K.W.M. holds the Adam J. Burgasser Chair in Astrophysics and is supported by an NSF Grant (2008031). 
A.B.P. is a Banting Fellow, McGill Space Institute (MSI) Fellow, and a Fonds de Recherche du Quebec -- Nature et Technologies (FRQNT) postdoctoral fellow. 
Z.P. is a Dunlap Fellow.
K.S. is supported by the NSF Graduate Research Fellowship Program. 
FRB research at UBC is supported by an NSERC Discovery Grant and by the Canadian Institute for Advanced Research. The CHIME/FRB baseband system is funded in part by a CFI JELF award to I.H.S. 
D.C.S is supported by by NSERC grant number RGPIN/03985-2021.
%

\facility{CHIME/FRB}
\software{Astropy \citep{2013A&A...558A..33A, 2018AJ....156..123A},
  Numpy \citep{harris2020array}, Scipy \citep{2020SciPy-NMeth}, 
  Matplotlib \citep{Hunter:2007}}

\begin{appendix}
\input{appendix}
\end{appendix}

\bibliography{main}{}
\bibliographystyle{aasjournal}
\end{document}

%% file: appendix.tex

\section{Other two methods of localization}\label{app: Other two methods of localization}

\subsection{Method 2}\label{subsec: method-2}

\begin{figure*}
\centering
\includegraphics[width=0.43\textwidth]{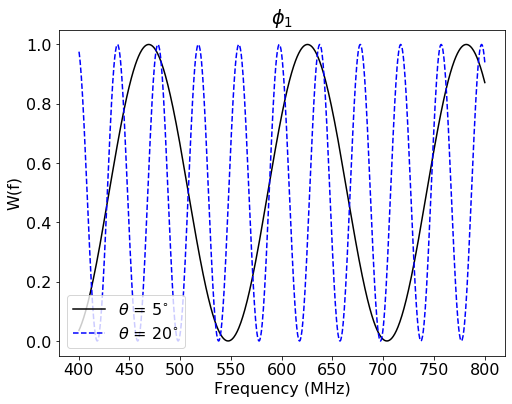}
\caption{Plots of the function W(f, $\theta$). Up: We assume the a flat power-law index of zero and plot the W(f, $\theta$) of the second beam column (i.e., $\phi_{1}$=0 deg at $\theta$ of 5 and 20 deg, respectively. 
\label{fig:Wf_panels}}
\end{figure*}

In the E-W direction, CHIME is an interferometer with 4 elements, corresponding to the focal lines of the 4 cylinders. The resulting beam response is thus an interference pattern of those 4 `slits' and can be thought of as analogous to double-slit interference. We therefore utilize the idea of double-slit interference\footnote{We revised the formula in Section 14.7 in \href{http://web.mit.edu/8.02t/www/802TEAL3D/visualizations/coursenotes/modules/guide14.pdf}{MIT course notes accessed on 22/Jun/2023}.} for testing the localization of the far side-lobe events, 
\begin{equation}\label{equation: freq_localization}
\mathrm{W}(f, \theta) = \sum\limits_{b=0}^{3} \mathrm{I}(f)\;\cos^{2}{(\frac{\pi{d}f}{c}\sin(\theta+\phi_{b}))}, 
\end{equation}
where I($f$) is the spectral profile of 1024 frequency channels of the brightest time-bin after masking any radio-frequency-interference (RFI) in the dedispersed dynamic spectrum of each beam, $d$ is 22 m for the separation of the CHIME cylinder focal lines \citep{2022ApJS..261...29A}, $f$ is the frequency in MHz, $c$ is the speed of light, $\theta$ is the position offset from meridian from -90 to +90 deg, $\phi_{b}$ accounts for the beamforming offset between the four beam columns such that $\phi_{0,1,2,3}$= $(-0.4, 0, 0.4, 0.8)$ deg, where the second beam is located on the meridian, and the others have an offset of 0.4 deg relative to the adjacent one. We assume that the I(f) is a flat spectrum with power-law index of zero, 

Figure \ref{fig:Wf_panels} shows W(f,$\theta$) at different $\theta$ angles. The separation of the spiky pattern across the frequency channels in one beam is related to the position offset, where the larger position offset yields a smaller separation of the spiky pattern. For instance, the spectral separation of spikes in a single beam with an offset of 5 and 20 deg is 156 and 40 MHz, respectively. The shift in the spiky pattern between beams in the 4 beam row, from East to West, tells us whether the source is in the Eastern or Western sky.

\begin{figure*}
\centering
\includegraphics[width=0.43\textwidth]{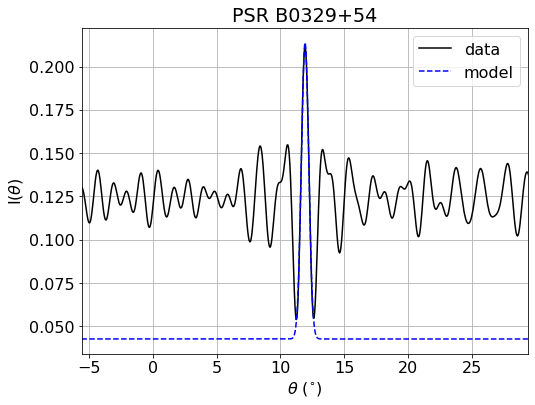}
\includegraphics[width=0.43\textwidth]{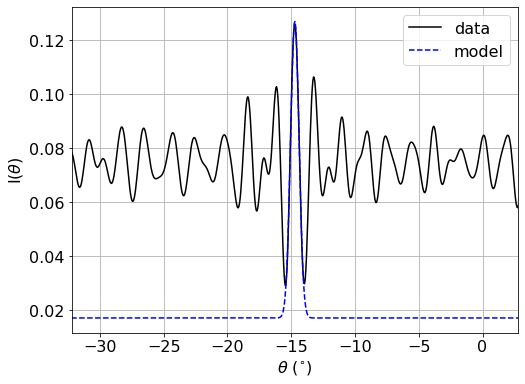}
\caption{I(f) of the far side-lobe event of B0329+54 (up) and B0531+21 (bottom). The solid and dashed lines represent the data and the Gaussian model, respectively. 
\label{fig:If_panels}}
\end{figure*}

We test the localization of the far side-lobe events of PSR B0329+54 and B0531+21, where their waterfalls are shown in Figure \ref{fig:waterfalls_ten_sidelobe}, with the following procedures. First, we mask the dynamic spectrum and convert the 1024 frequency channels of the brightest time-bin into the spectral profile, W($f$).  Second, we sum up the W($\theta, f$) of the four E-W beams and average over the frequency into I($\theta$). Third, we fit a Gaussian profile to I($\theta$) at its maximal peak and determine the best $\theta$ and the 68$\%$ confidence interval. Figure \ref{fig:If_panels} shows the I($\theta$) and the Gaussian fit for the far side-lobe events of PSR B0329+54 and B0531+21. Last, we assume the source is on the trajectory of the second Western beam, which points to the meridian, and we find the best localization corresponding to $\theta$. For this method, the localization offset is $\sim$1 deg for the HA from 15 to 50 deg as shown in Figure \ref{fig:two_comparisons}. 

\subsection{Method 3} \label{subsec: method-3}
A third complementary far side-lobe localization method takes advantage of the distinct, knotty spectrum of a far side-lobe event detected by inteferometrically combining signals from multiple detectors. The separation of these spiky patches of the spectrum is correlated to the degree of offset from the meridian. We inject simulated Gaussian bursts at a wide range of meridian offsets to empirically fit for this correlation, and found that the relation between the knot separation ($y$) in MHz and the East-West offset from Meridian ($x$) in degree can be represented by: 
\begin{equation} 
y = 772.943/x + 1.844\,.
\label{eq:m3}
\end{equation} 
We can see that the closer the separation and hence the larger number of knots in the spectrum, the further out the source is in the side lobe. On the other hand, sources within $\sim$3 deg of the meridian do not show any knotty spectrum as they are within the main lobe. 
Similar to what was described in Section~\ref{subsec: method-2}, we can tell the direction (East or West of the meridian) of the source by comparing the spectrum across the four East-West beams of the same row. In principle, this method is less susceptible to Radio Frequency Interference (RFI) present in part of the spectrum, because we only need to measure one of the knot separations to model the corresponding E-W offset. In practice, the fact that the intensity data has limited spectral resolution introduces uncertainty in the fitted knot separation, 
a feature we attempt to overcome by averaging as many knot separations observed for each burst. Systematic offsets can be seen in the coordinates determined by this method in Fig.~{\ref{fig:three_comparisons}. This is most likely because simulated signals were only generated for the meridian at zenith angle = 0 $\deg$, leading to error in the empirical relationship in Equation~\ref{eq:m3}.
}

\subsection{Comparisons}

\begin{figure*}
\centering
\includegraphics[width=0.47\textwidth]{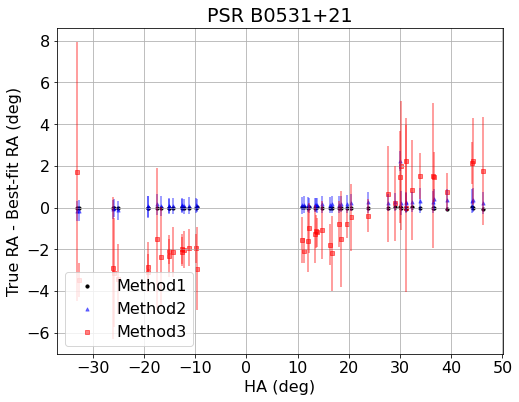}
\includegraphics[width=0.47\textwidth]{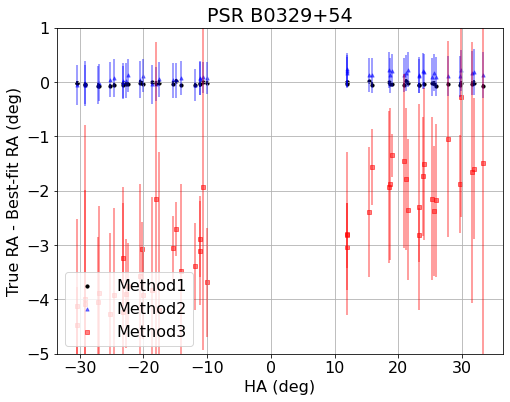}
\includegraphics[width=0.47\textwidth]{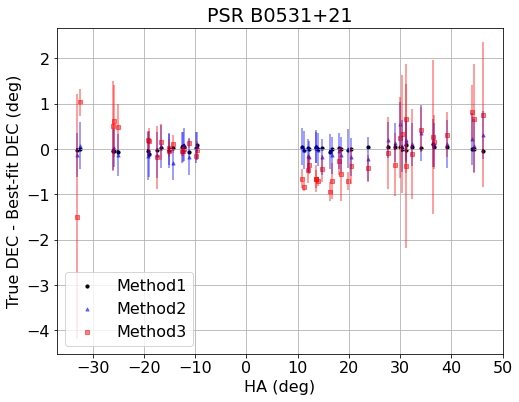}
\includegraphics[width=0.47\textwidth]{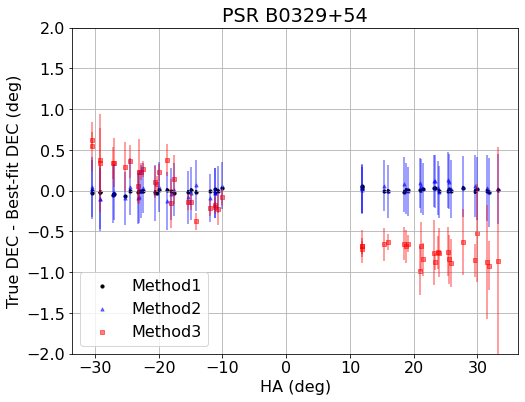}
\caption{The localization offset comparisons of three methods. We note that Method 1 has the smallest systematic errors and Method 3 has the largest ones.
\label{fig:three_comparisons}}
\end{figure*}

\begin{figure*}
\centering
\includegraphics[width=0.47\textwidth]{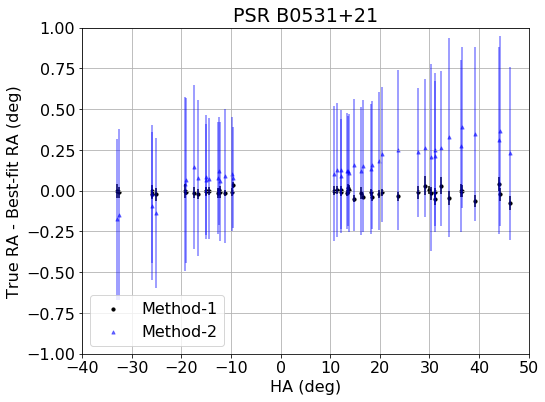}
\includegraphics[width=0.47\textwidth]{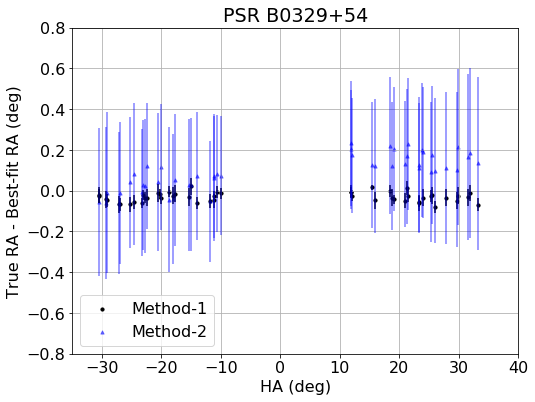}
\includegraphics[width=0.47\textwidth]{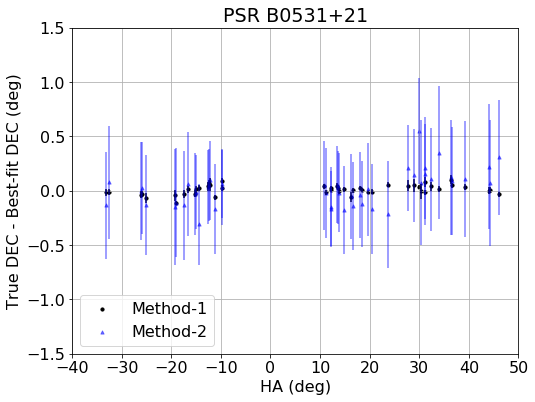}
\includegraphics[width=0.47\textwidth]{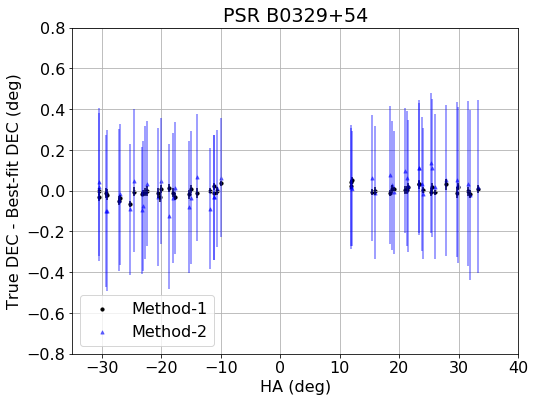}
\caption{The localization offset comparisons of Method 1 and Method 2. We did not understand the origin of the systematic errors of Method 2, and the localization with voltage data in the future could further improve the localization precision.
\label{fig:two_comparisons}}
\end{figure*}

Figure \ref{fig:three_comparisons} shows the localization comparisons for the three methods. Figure \ref{fig:two_comparisons} shows the localization comparison for Method 1 and Method 2. For the three methods, we note that Method 1 has the smallest systematic error, and Method 3 has the largest one. We did not understand the origin of the systematic error. Further investigation with the voltage data of far side-lobe events from pulsars may be helpful to better understand the systematic errors.

\section{Diagnostic plots for the 10 far side-lobe events}\label{app: Waterfalls of ten far side-lobe events}

\begin{figure*}
\centering
\includegraphics[width=0.47\textwidth]{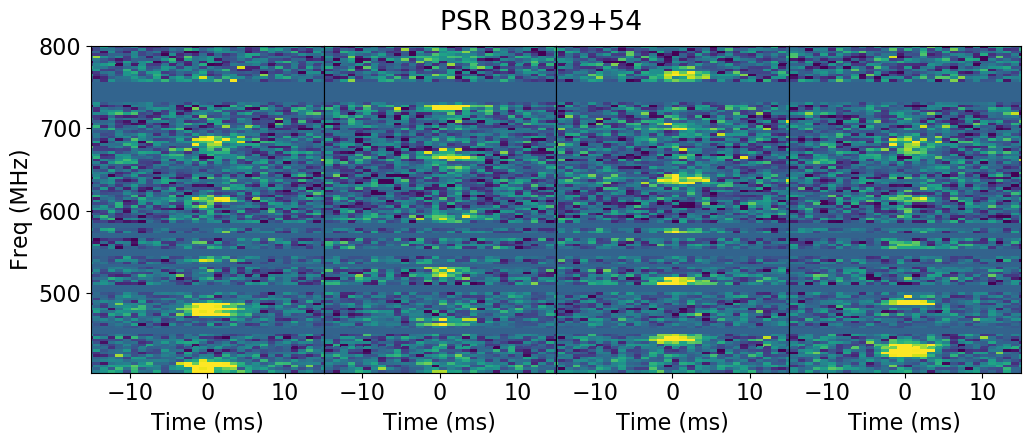}
\includegraphics[width=0.47\textwidth]{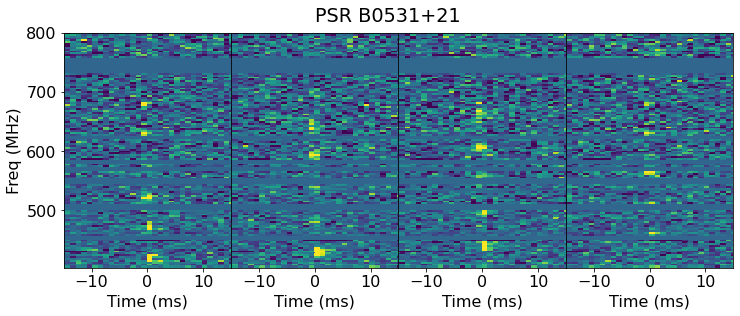}
\includegraphics[width=0.47\textwidth]{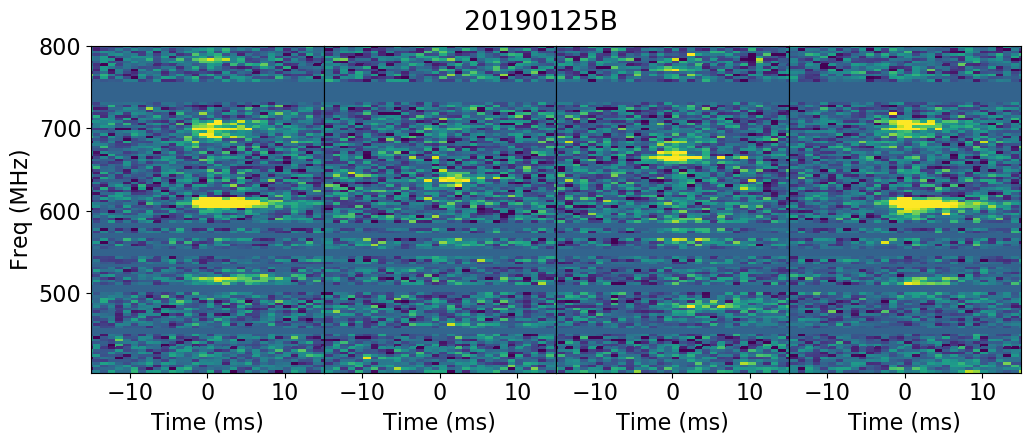}
\includegraphics[width=0.47\textwidth]{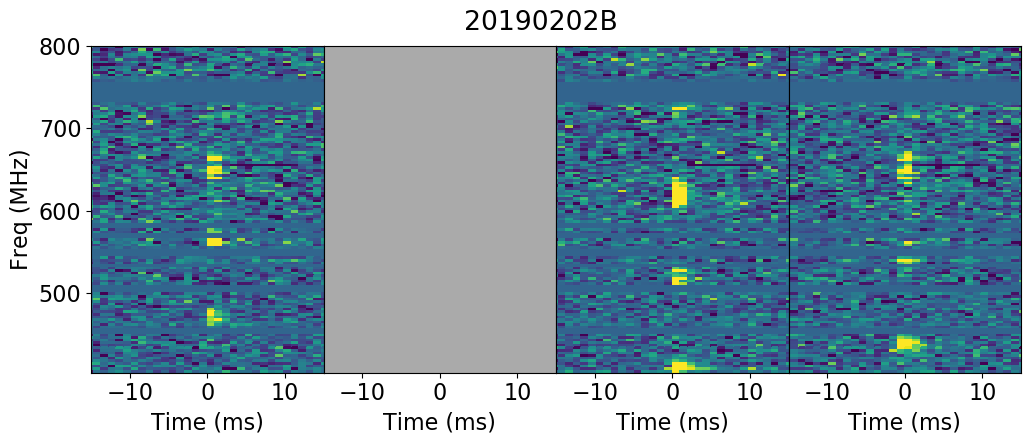}
\includegraphics[width=0.47\textwidth]{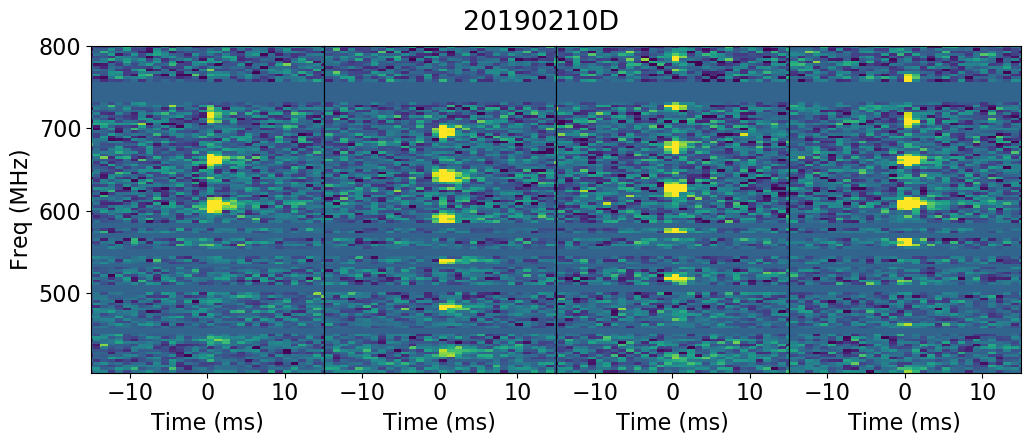}
\includegraphics[width=0.47\textwidth]{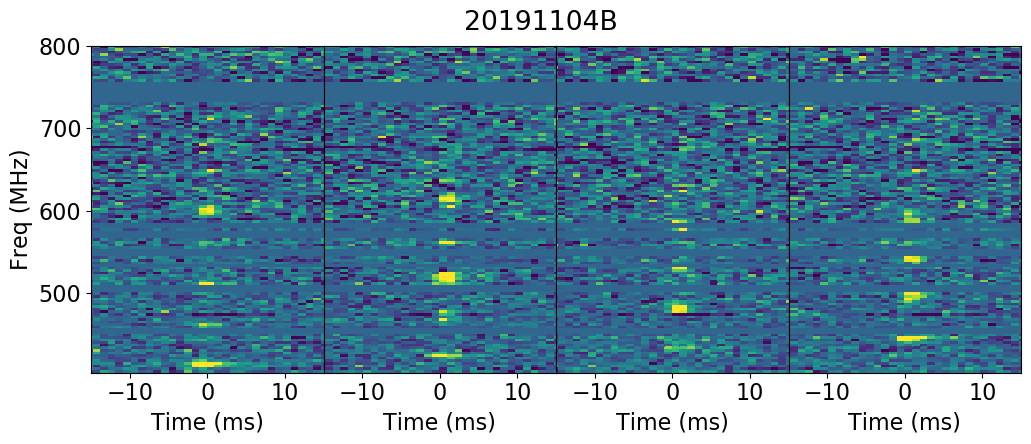}
\includegraphics[width=0.47\textwidth]{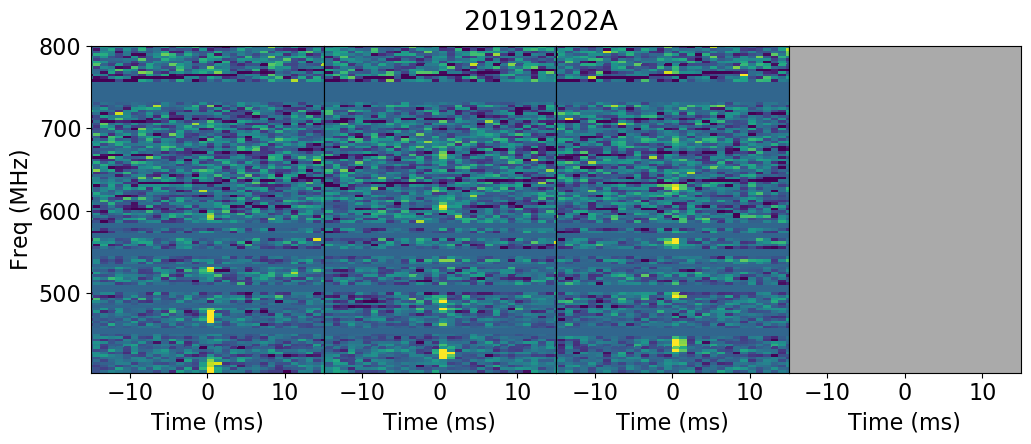}
\includegraphics[width=0.47\textwidth]{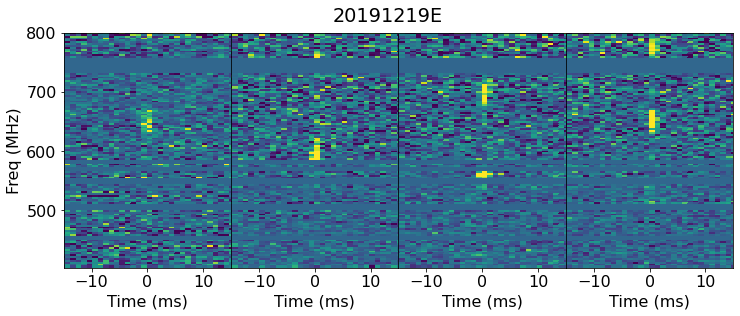}
\includegraphics[width=0.47\textwidth]{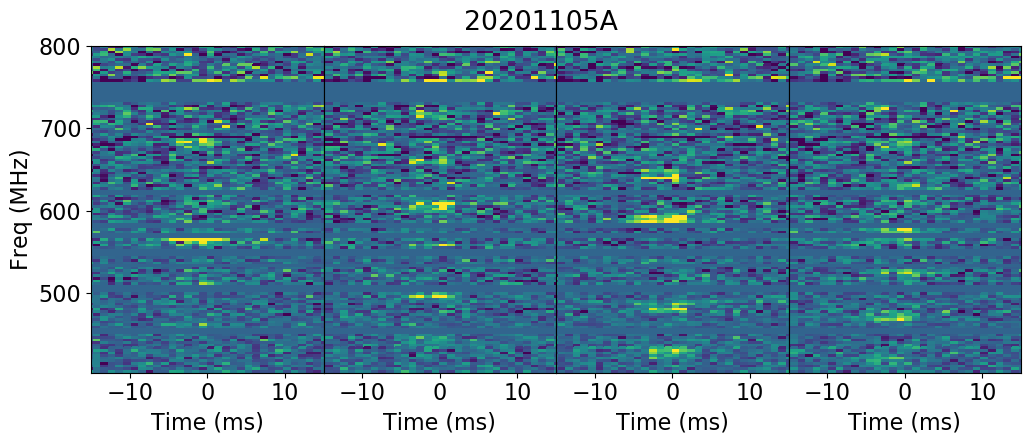}
\includegraphics[width=0.47\textwidth]{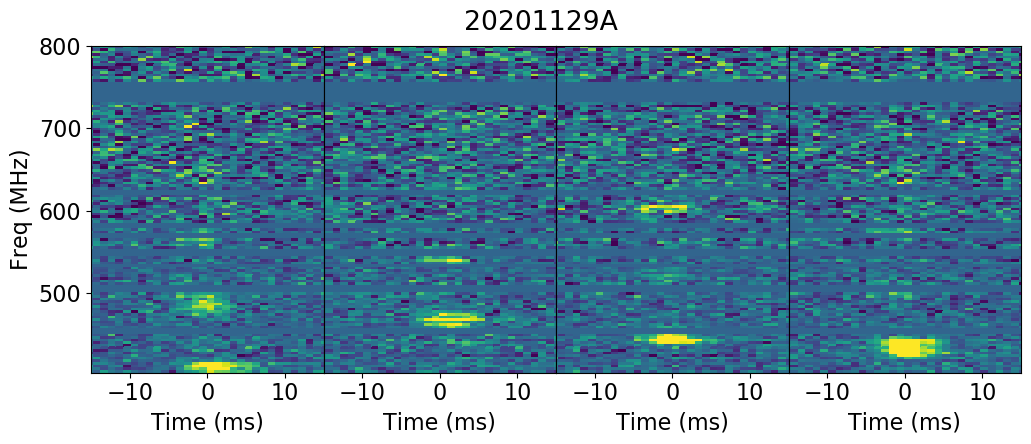}
\includegraphics[width=0.47\textwidth]{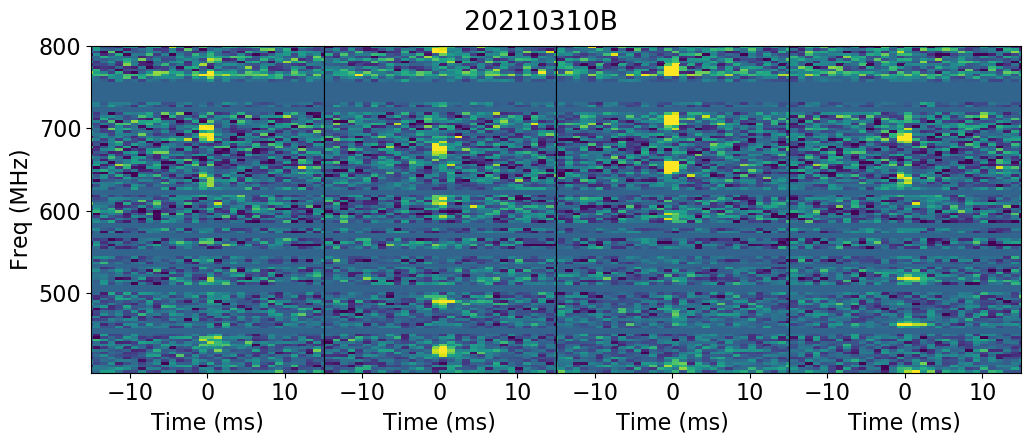}
\includegraphics[width=0.47\textwidth]{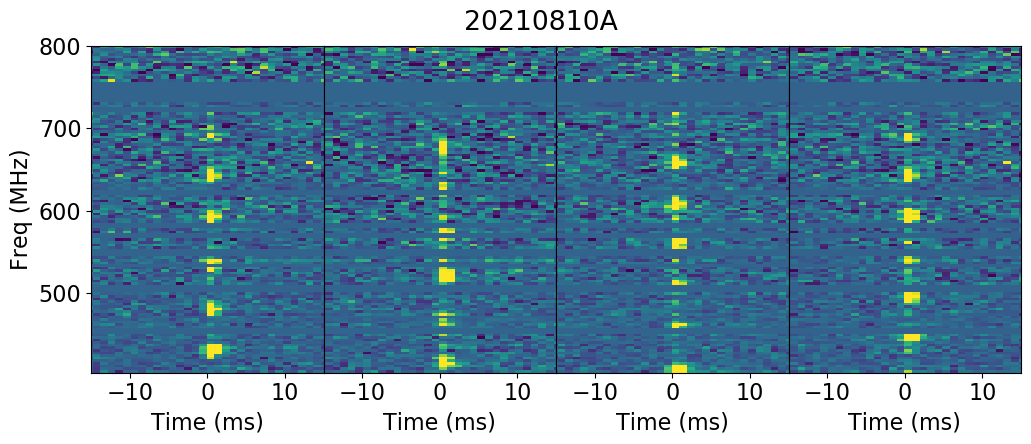}
\caption{The waterfalls of the far side-lobe events, including one from PSR B0329+54, another from PSR B0531+21, and the others from FRBs. Each panel represents for one far side-lobe event. The sub-panel, the timing and frequency resolutions are described in Figure \ref{figure:1}. For events 20190202B, 20191202A, and 201912E, only three of four W-E beams in the same N-S row are above the detection threshold. For events 20190125B, 20191104B, and 20210810A, there are lower-S/N detections in adjacent beam rows, which are not shown.
\label{fig:waterfalls_ten_sidelobe}}
\end{figure*}

\begin{figure*}
\centering
\includegraphics[width=0.47\textwidth]{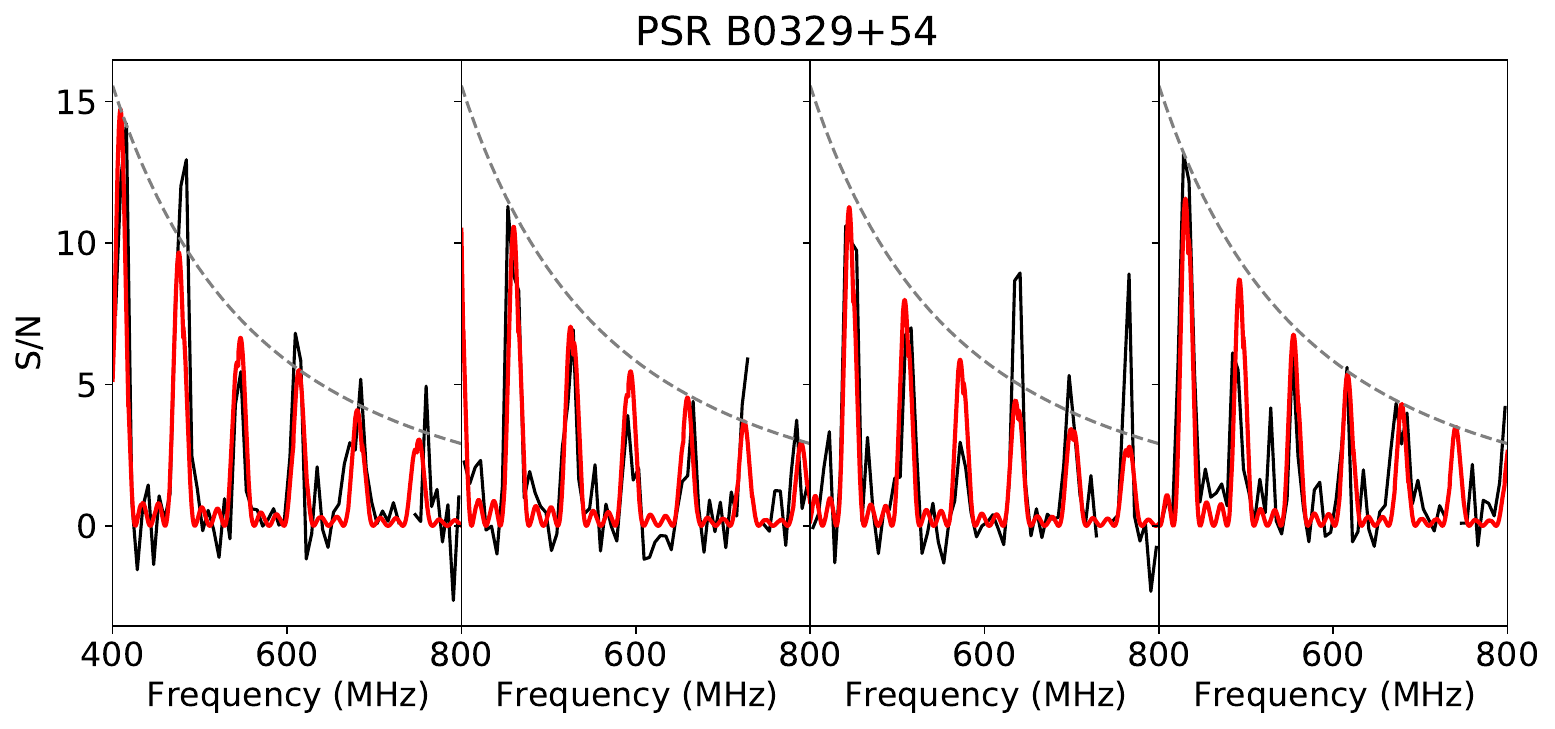}
\includegraphics[width=0.47\textwidth]{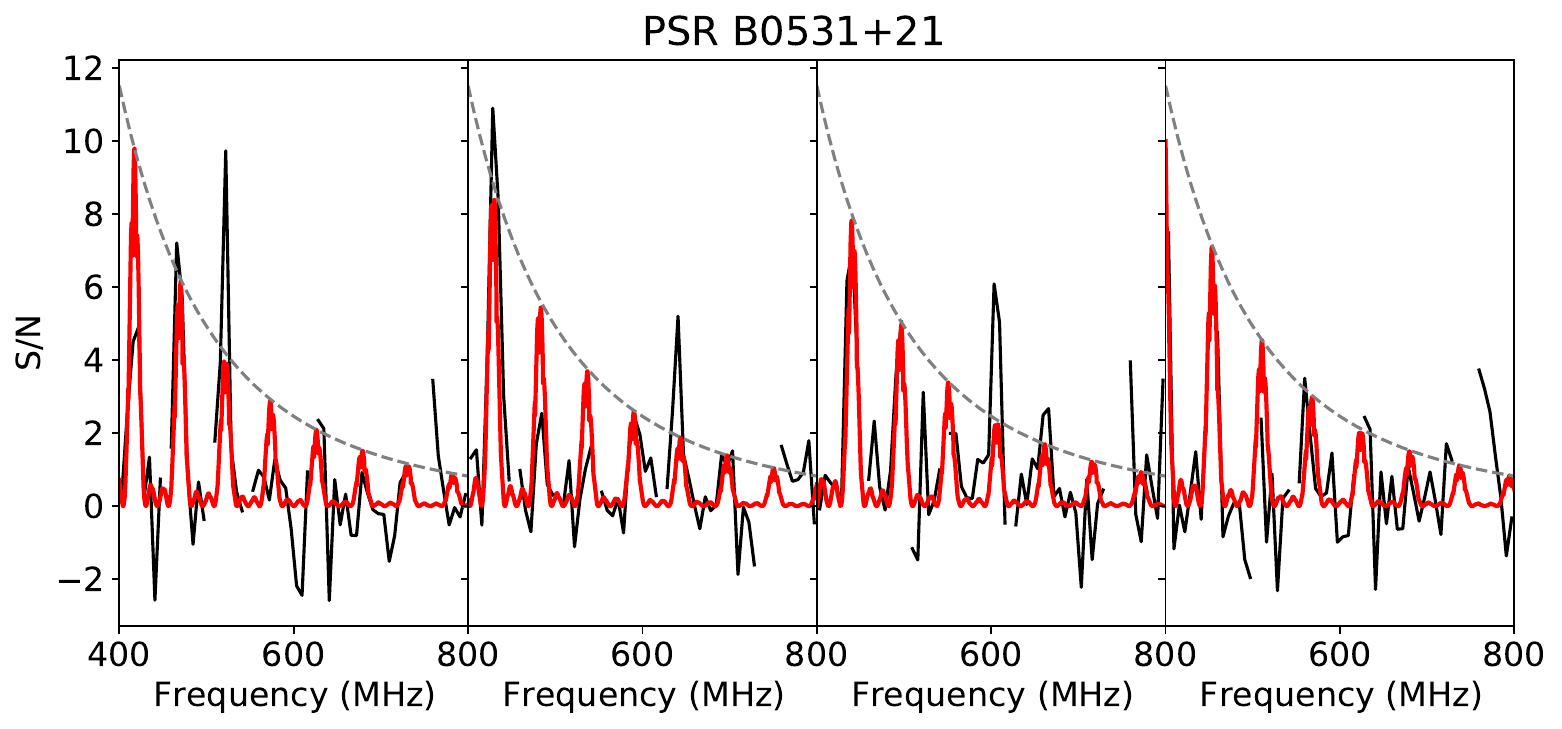}
\includegraphics[width=0.47\textwidth]{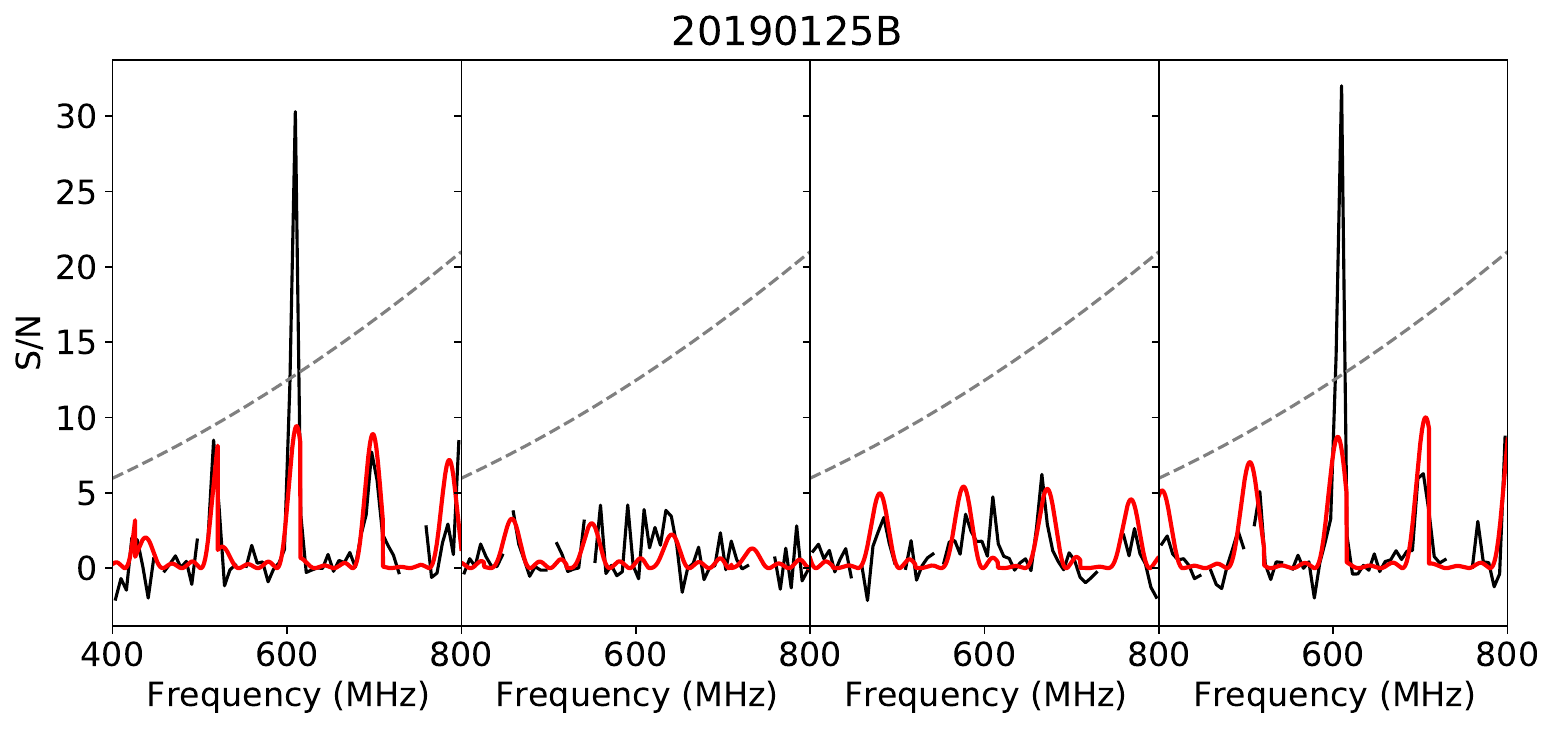}
\includegraphics[width=0.47\textwidth]{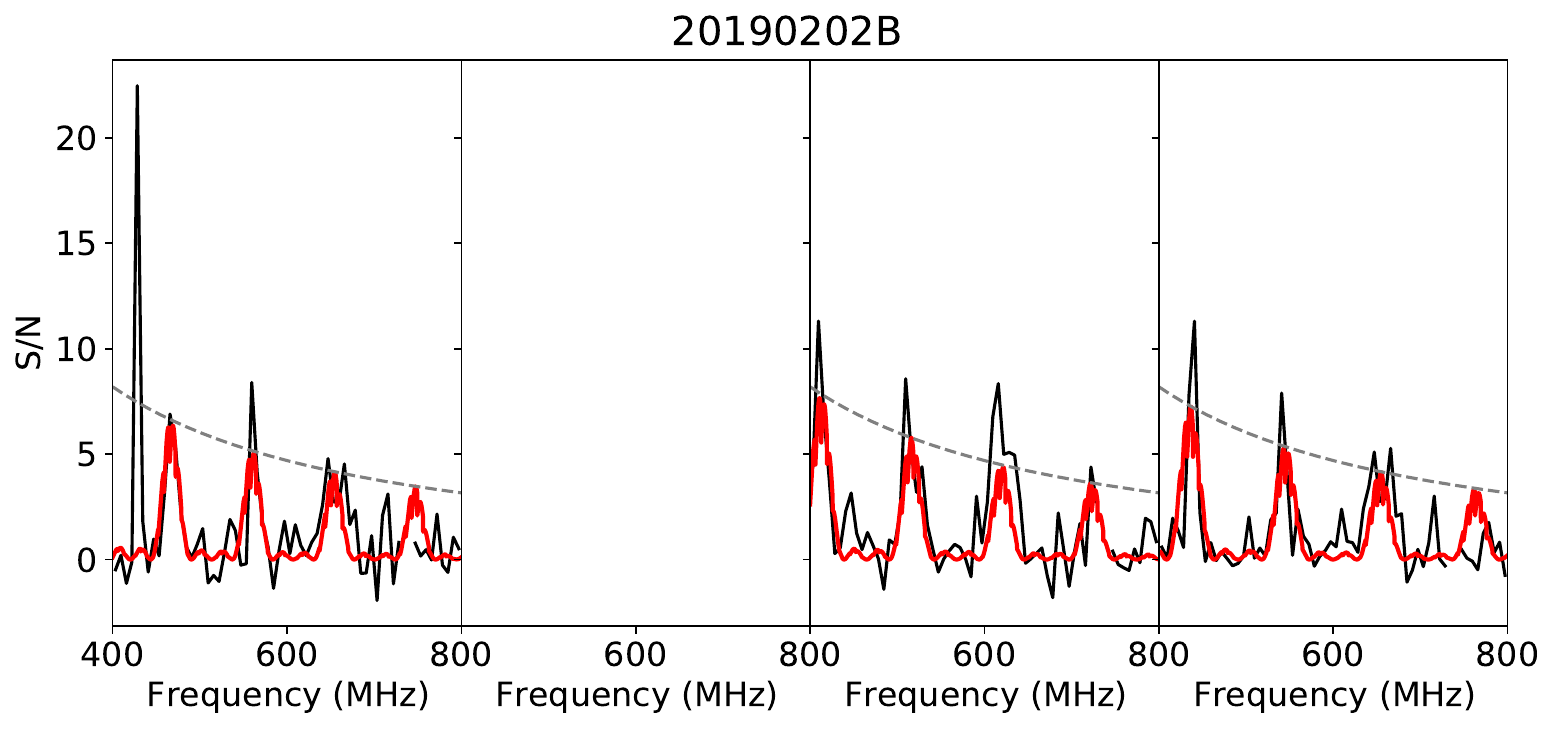}
\includegraphics[width=0.47\textwidth]{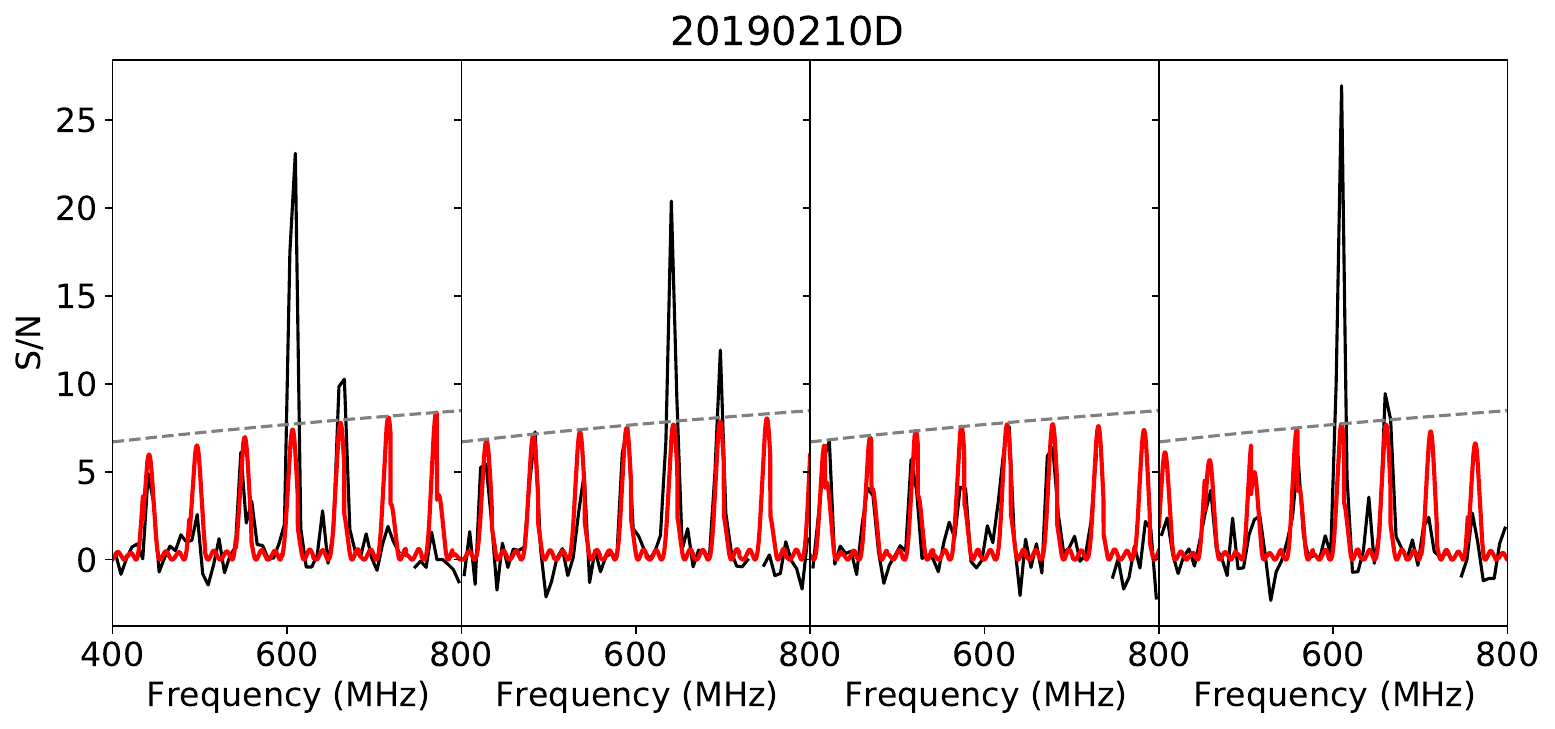}
\includegraphics[width=0.47\textwidth]{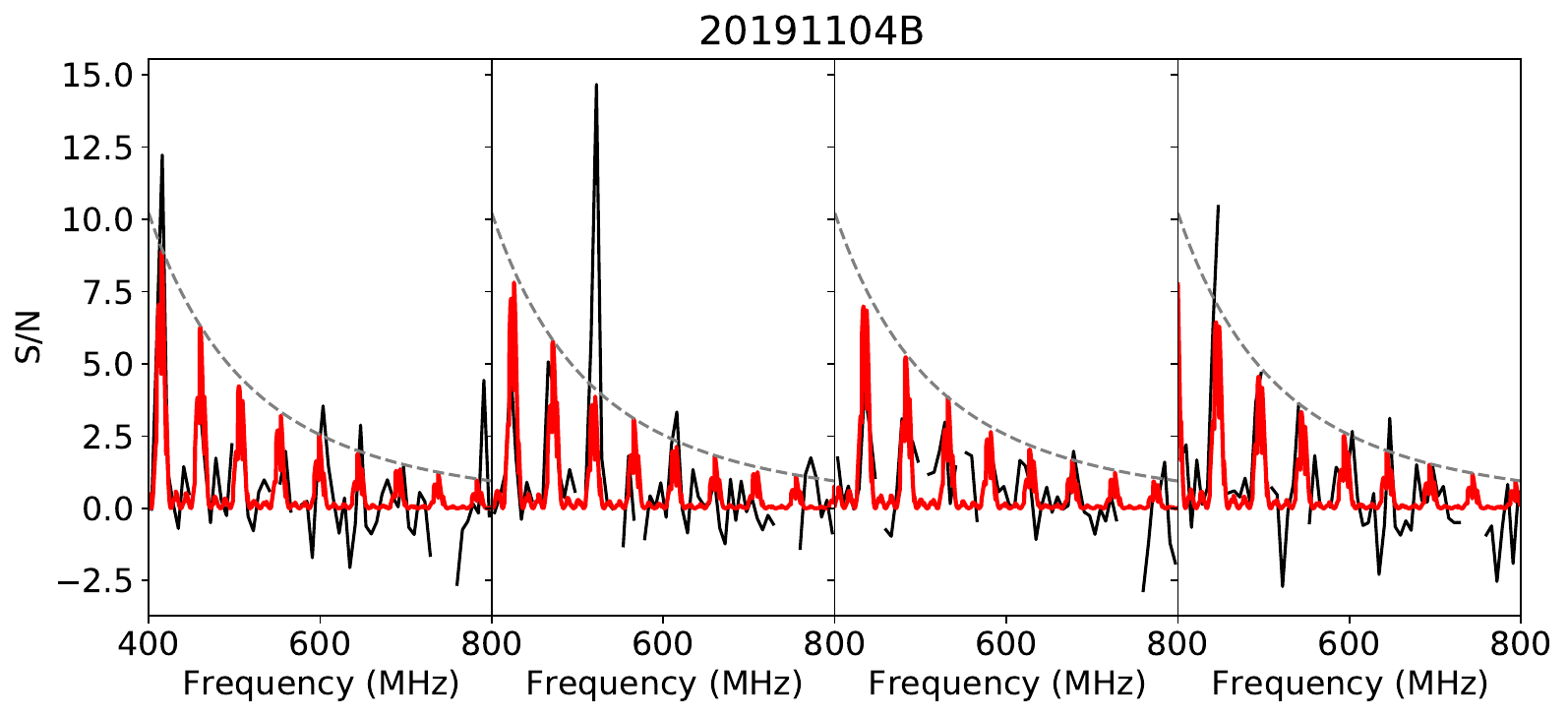}
\includegraphics[width=0.47\textwidth]{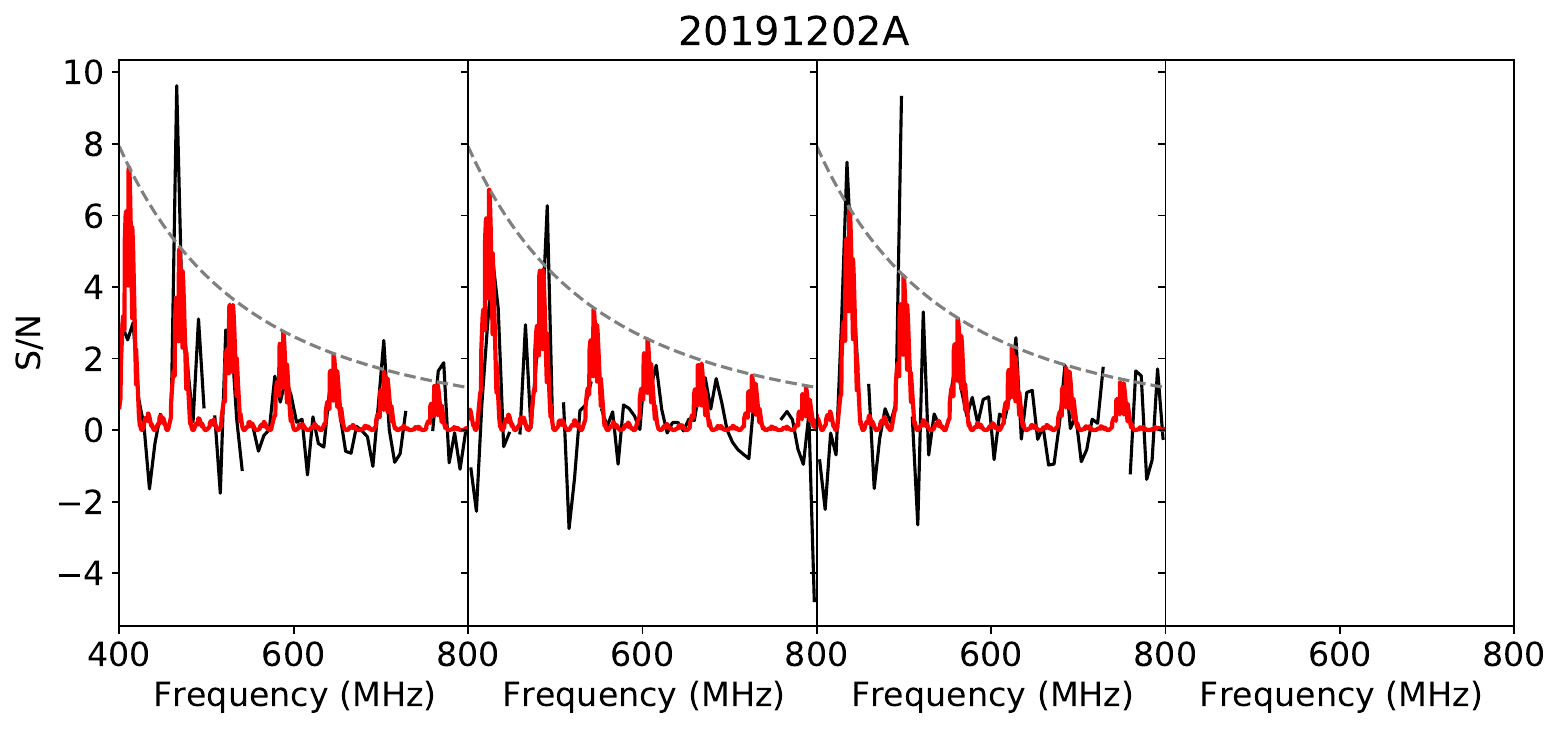}
\includegraphics[width=0.47\textwidth]{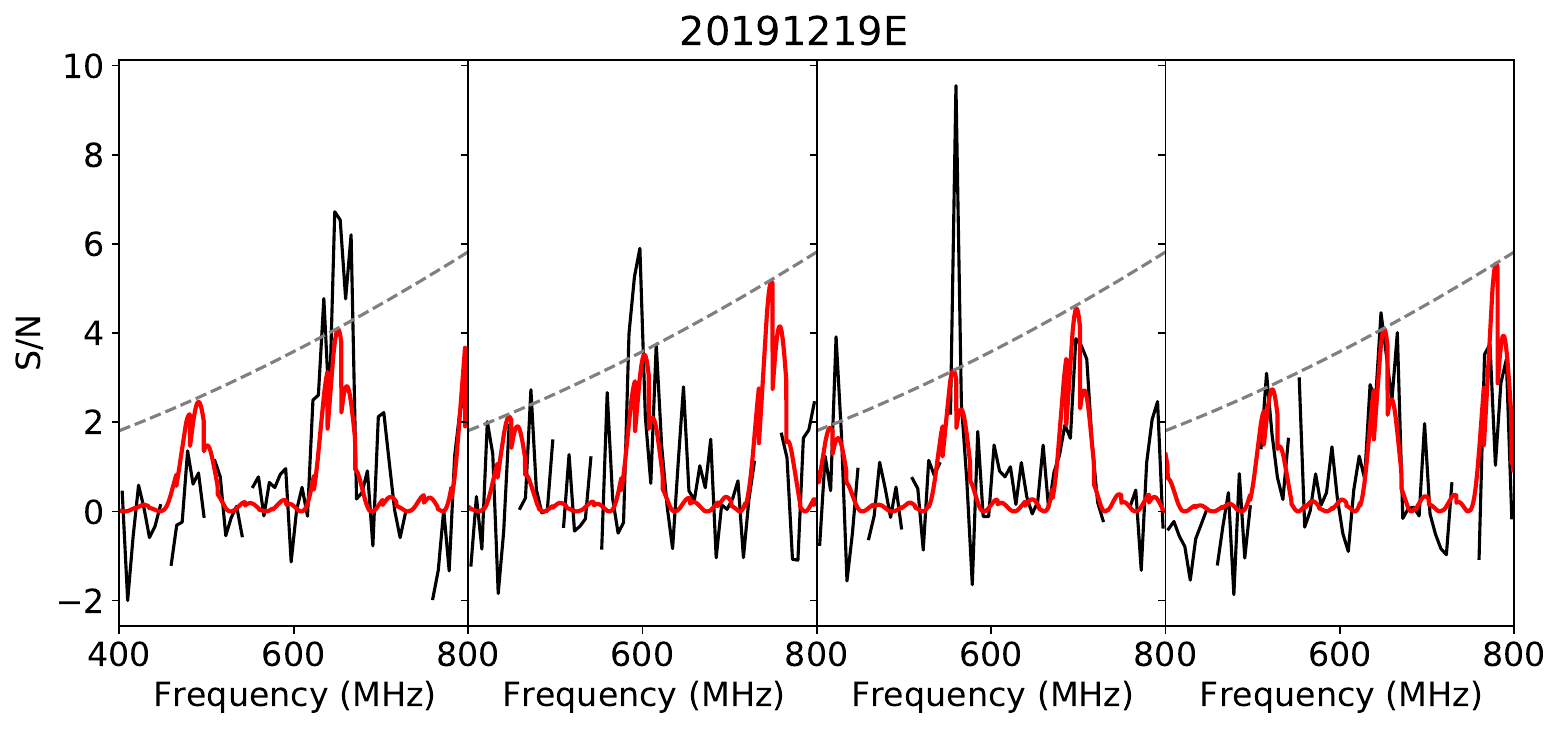}
\includegraphics[width=0.47\textwidth]{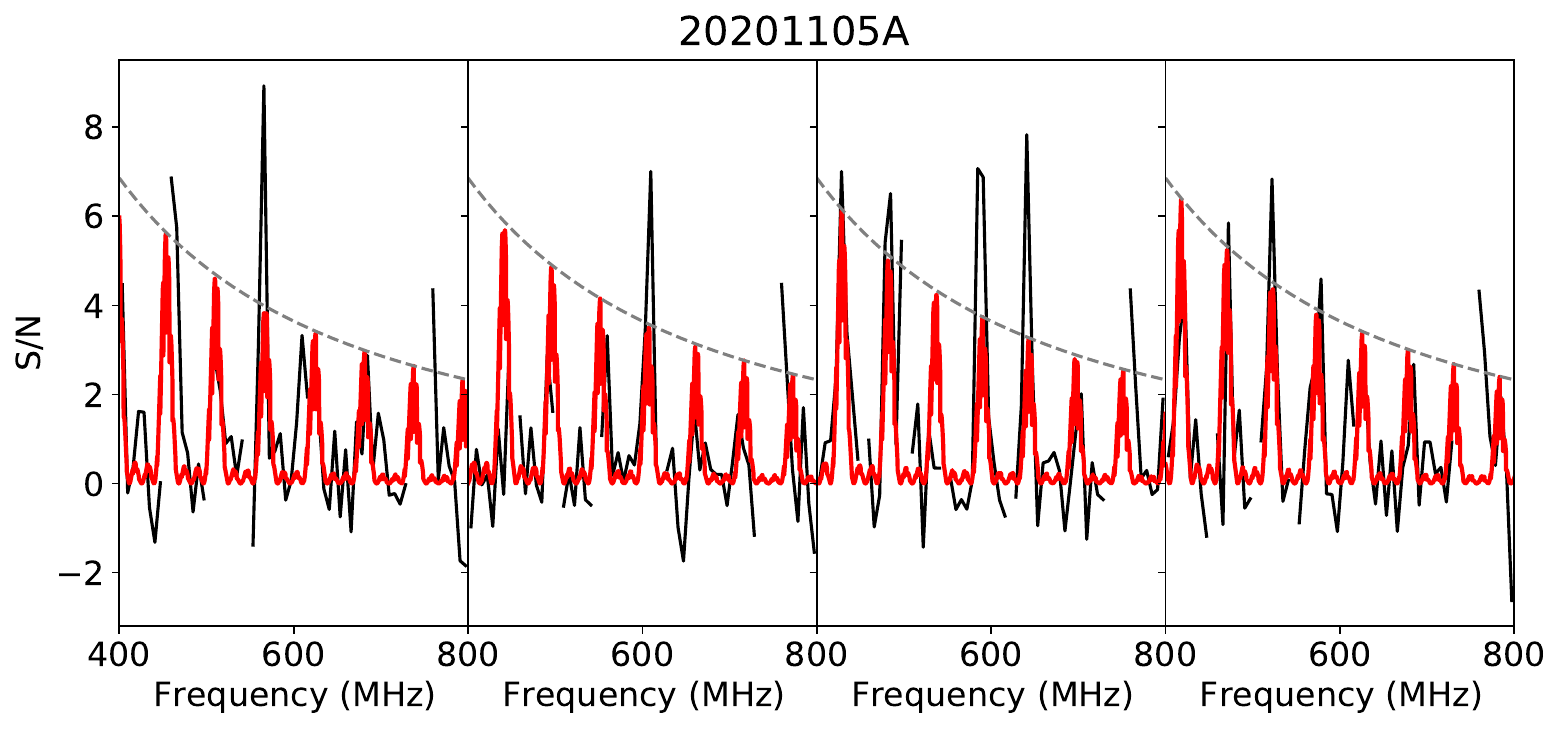}
\includegraphics[width=0.47\textwidth]{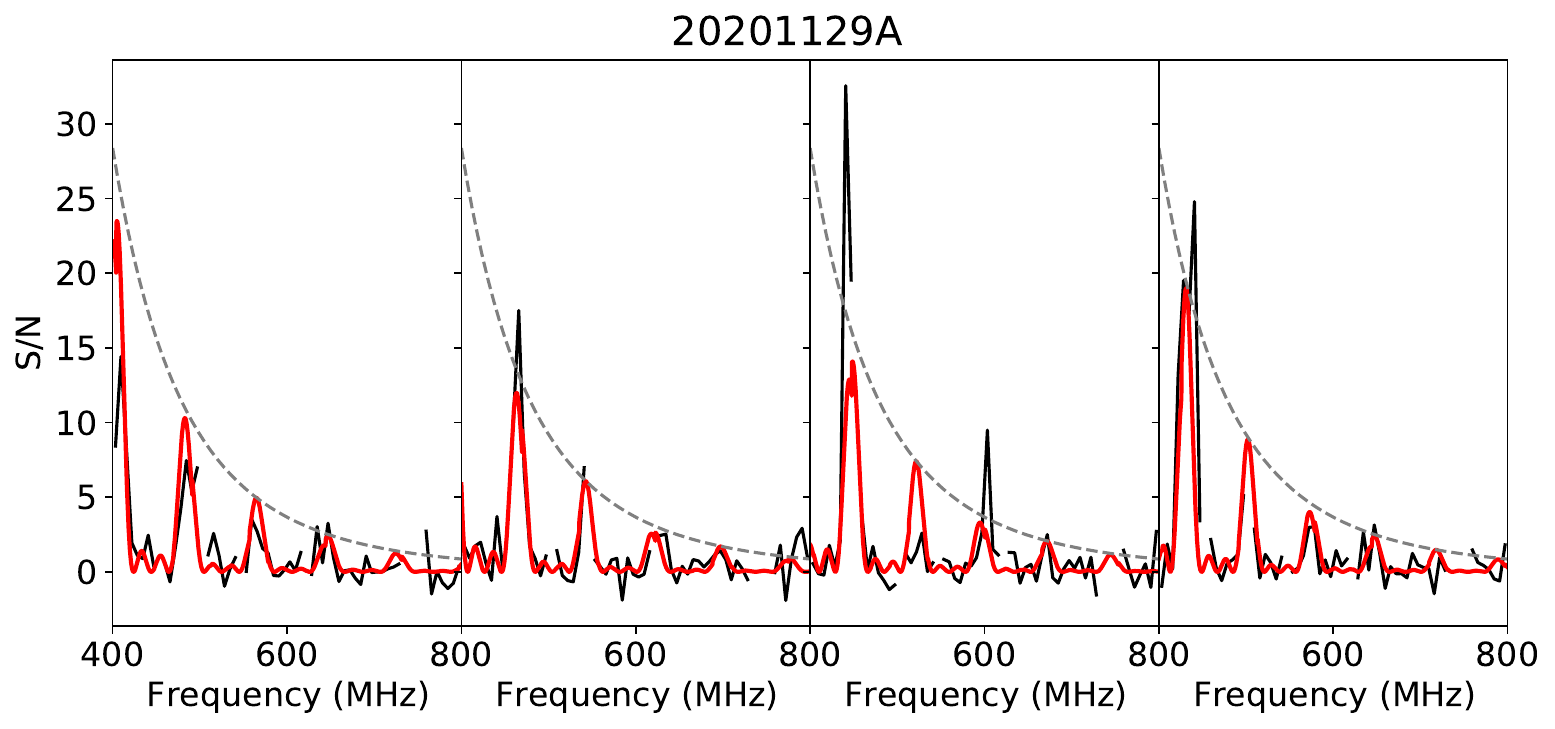}
\includegraphics[width=0.47\textwidth]{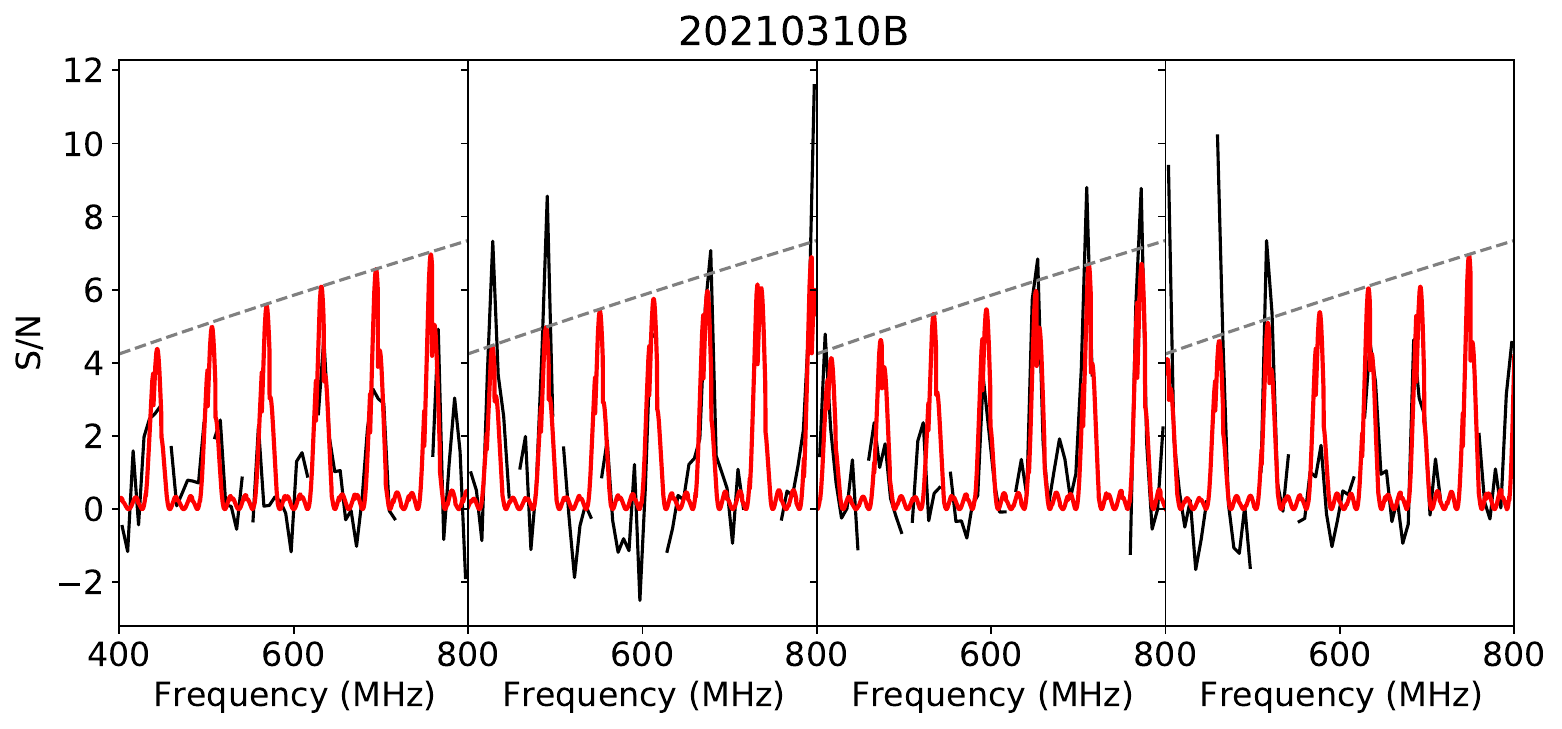}
\includegraphics[width=0.47\textwidth]{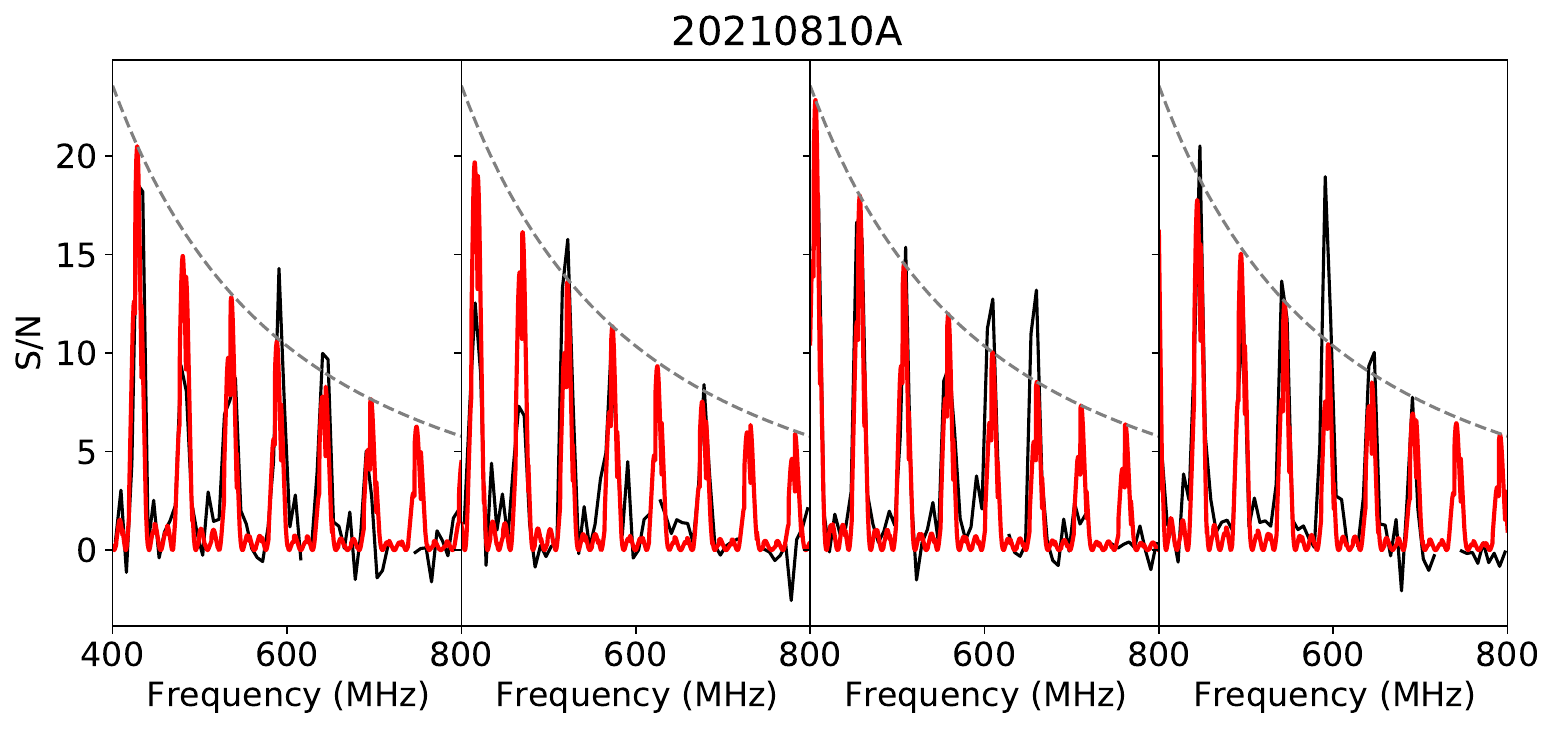}
\caption{
Spectra (solid black) and fitted model (solid red) for the Method 1 intensity localizations of example pulses from PSRs B0329+51 and B0531+21 and the 10 side-lobe events. The underlying fitted Power-law model FRB spectrum is plotted as grey dashed lines.
\label{fig:spec_fits_ten_sidelobe}}
\end{figure*}

\begin{figure*}
\centering
\includegraphics[width=0.47\textwidth]{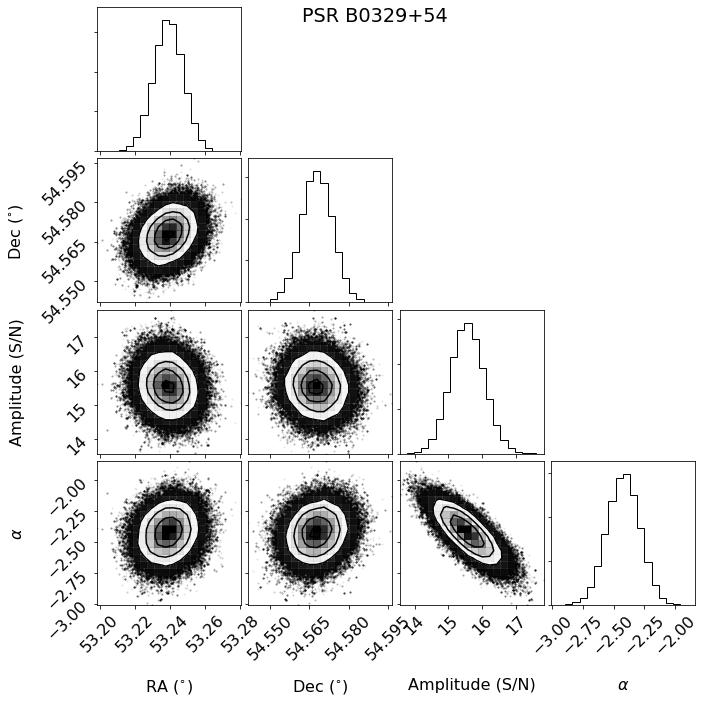}
\includegraphics[width=0.47\textwidth]{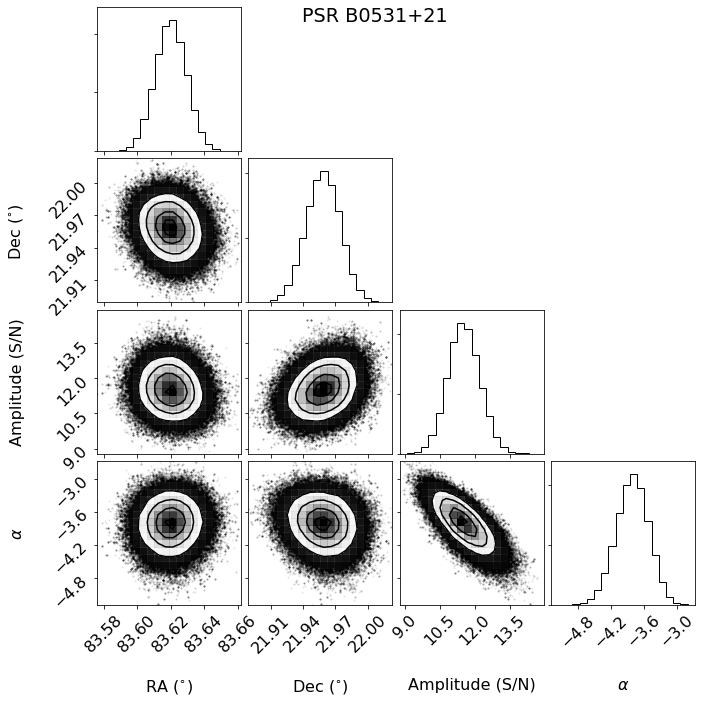}
\includegraphics[width=0.47\textwidth]{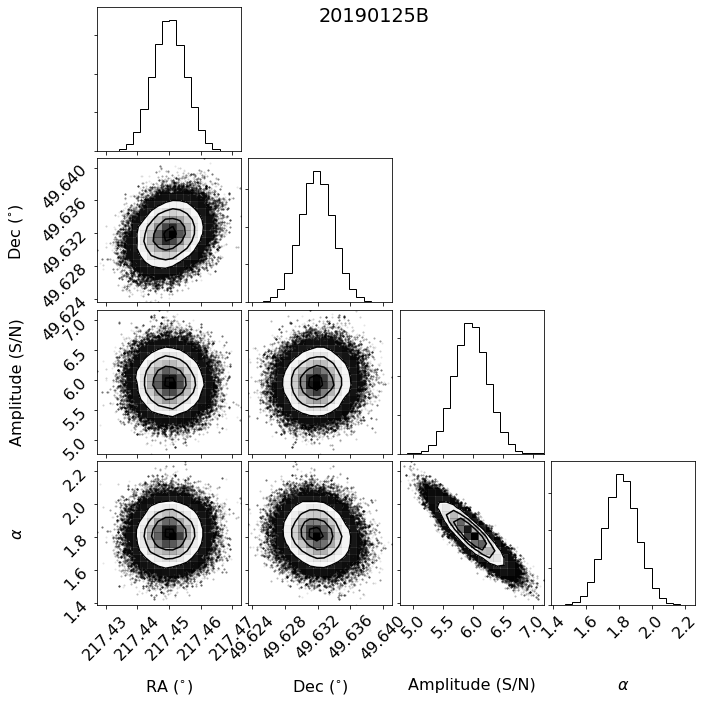}
\includegraphics[width=0.47\textwidth]{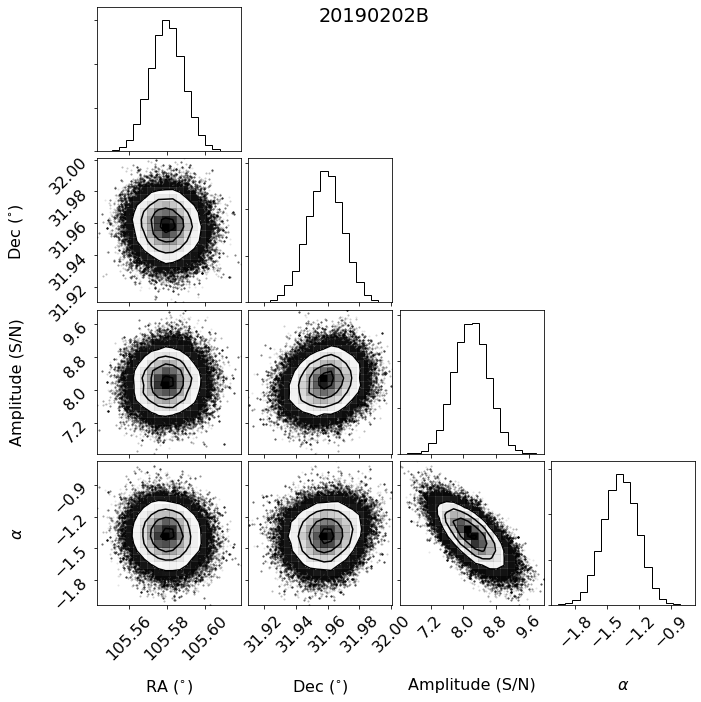}
\caption{Distributions of the posterior samples from the Method 1 intensity localizations of sample pulses from PSRs B0329+54 and B0531+21 and for the side-lobe FRBs~20190125B and 20190202B. 
\label{fig:corner_plots_ten_sidelobe1}}
\end{figure*}

\begin{figure*}
\centering
\includegraphics[width=0.47\textwidth]{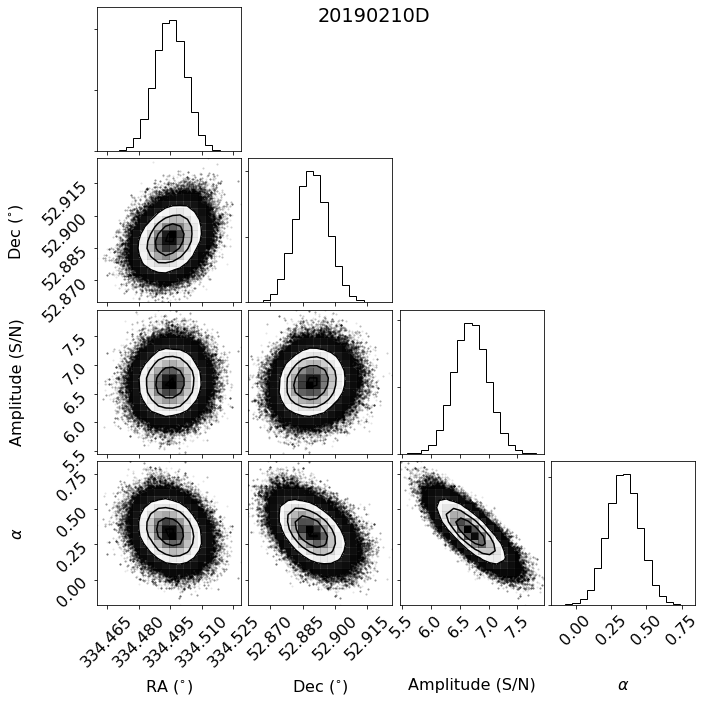}
\includegraphics[width=0.47\textwidth]{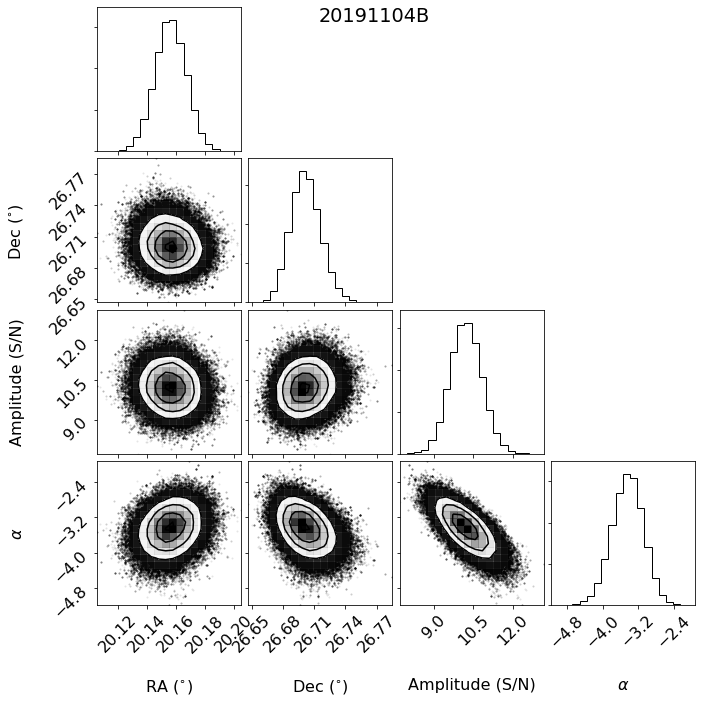}
\includegraphics[width=0.47\textwidth]{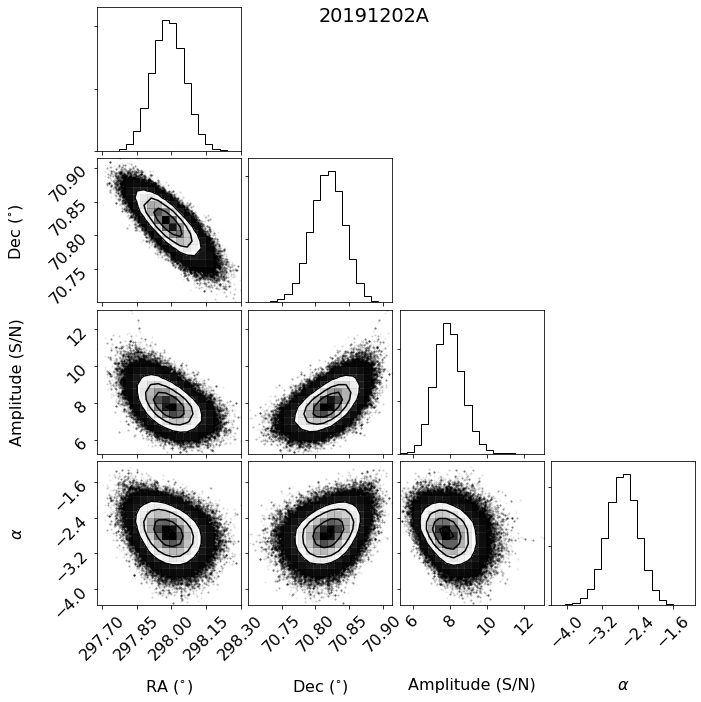}
\includegraphics[width=0.47\textwidth]{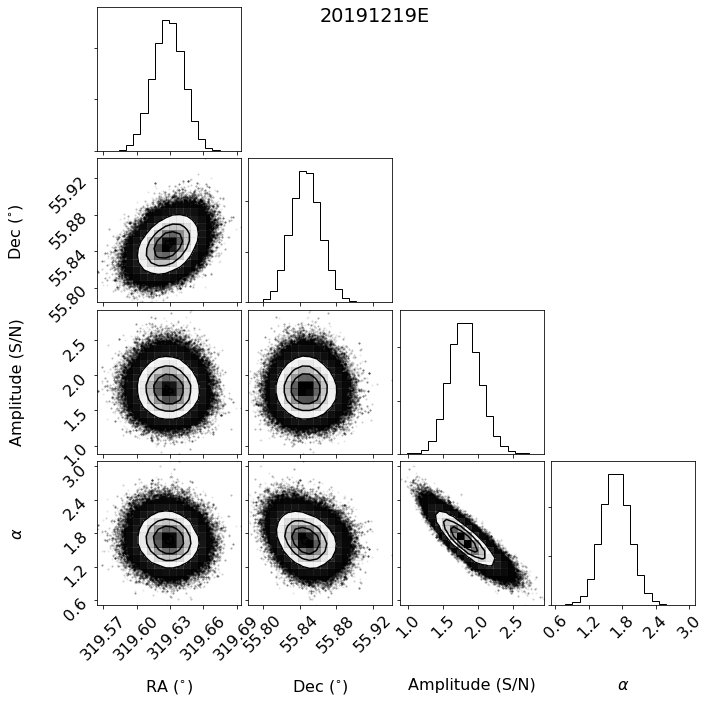}
\caption{Distributions of the posterior samples from the Method 1 intensity localizations of sample pulses for the side-lobe FRBs~20190210D, 20191104B, 20191202A, and 20191219E.
\label{fig:corner_plots_ten_sidelobe2}}
\end{figure*}

\begin{figure*}
\centering
\includegraphics[width=0.47\textwidth]{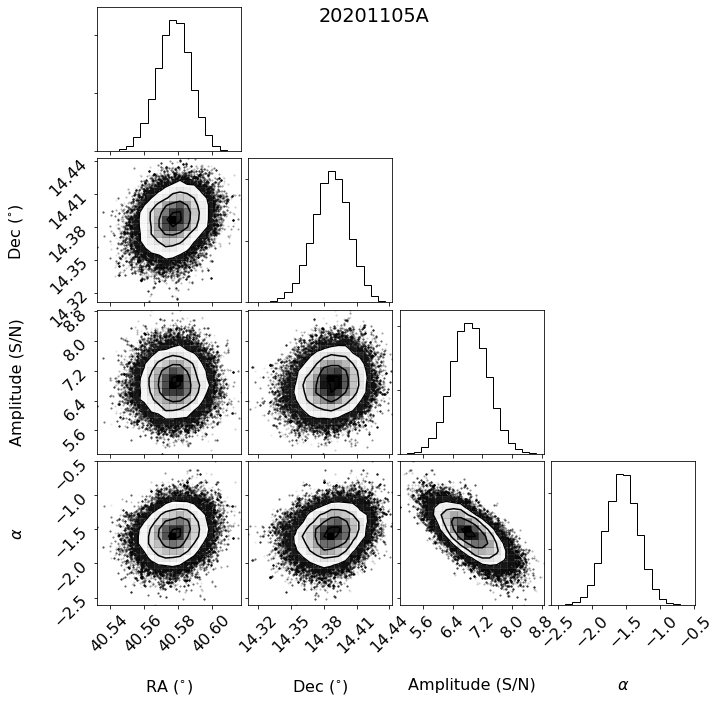}
\includegraphics[width=0.47\textwidth]{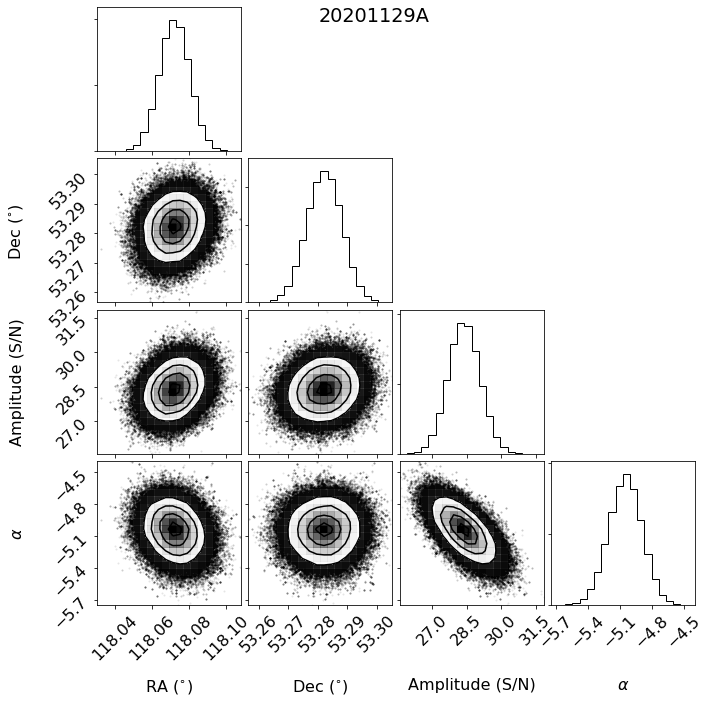}
\includegraphics[width=0.47\textwidth]{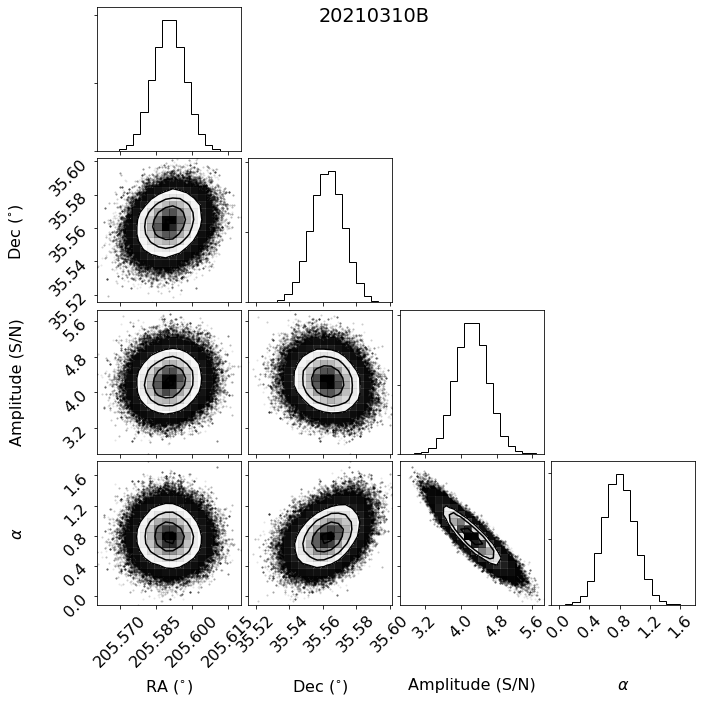}
\includegraphics[width=0.47\textwidth]{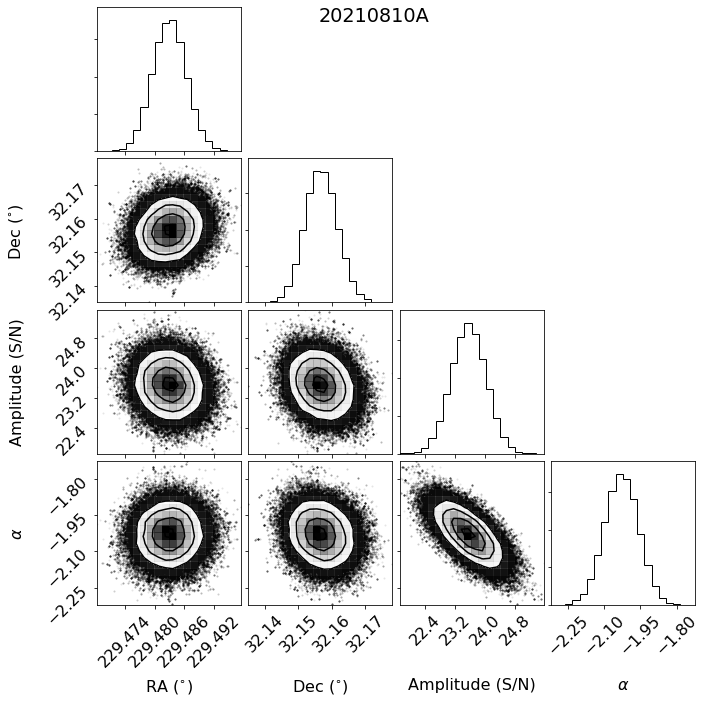}
\caption{Distributions of the posterior samples from the Method 1 intensity localizations for the side-lobe FRBs~20201105A, 20201129A, 20210310B, and 20210810A.
\label{fig:corner_plots_ten_sidelobe3}}
\end{figure*}

\psedit{Here we show the data sets used for localizations, the best-fit models, and the posterior distributions from the Method 1 MCMC sampling.}
Figure \ref{fig:waterfalls_ten_sidelobe} shows the dynamic spectra of all far side-lobe FRBs and an example of far side-lobe detections of B0329+54 and B0531+21. Figure \ref{fig:spec_fits_ten_sidelobe} shows the spectra and fitted model for the Method 1 intensity localizations of example pulses from PSRs B0329+51 and B0531+21 and the 10 side-lobe events. Figures \ref{fig:corner_plots_ten_sidelobe1}--\ref{fig:corner_plots_ten_sidelobe3} show the distributions of the posterior samples from the Method 1 intensity localizations for example pulses from PSRs B0329+51 and B0531+21 and the 10 side-lobe events.

\section{The SVD analysis of beam model}\label{app: the SVD analysis of beam model}

Appendix Table \ref{table: calibratos} shows the calibrators used for the holographic observations in Section \ref{section: Holographic calibration} to which the SVD decomposition of the beam model is applied. 

\begin{deluxetable*}{lccccccc}
  \tablecaption{The calibrators for the holography analysis.\label{table: calibratos}}
  \tablewidth{\textwidth}
  \tablehead{
    Source & RA\tablenotemark{a} & DEC\tablenotemark{b} & S($\nu$=725 MHz)\tablenotemark{c} & S($\nu$=600 MHz)\tablenotemark{d} & S($\nu$=400 MHz)\tablenotemark{e} & Date1 & Date2
    }
  \startdata
  CasA  & 23h23m27.94s   & +58d48m42.4s   & 2903.4$\pm$6.0 & 3343.8$\pm$6.8 & 4534.2$\pm$9.6  & 2018 Sep 29 & 2019 Dec 13 \\
  3C295 & 14h11m20.519s  & +52d12m09.97s  & 37.3$\pm$0.1   & 42.3$\pm$0.1   & 54.2$\pm$0.1    & 2019 Mar 02 & 2019 Mar 18 \\
  CygA  & 19h59m28.3566s & +40d44m02.096s & 3053.2$\pm$6.5 & 3626.1$\pm$8.1 & 5103.5$\pm$16.3 & 2018 Oct 17 & 2018 Oct 23 \\
  \enddata
  \tablenotetext{a-b}{\;\;\;\;\;The right ascension and the declination of the calibrator (J2000), taken from the NASA/IPAC Extragalactic Database \citep{2007AA...467..585B, 2008AJ....136.2373R, 2004AJ....127.3587F,  1971PASP...83..491M}.}
  \tablenotetext{c-e}{\;\;\;\;\;The flux density (Jy) is at 726, 600, and 400 MHz, which we infer with the polynomial function and parameters in \citet{2017ApJS..230....7P}.}
\end{deluxetable*}

\begin{figure*}[htp!]
  \centering
  \includegraphics[width=2\columnwidth]{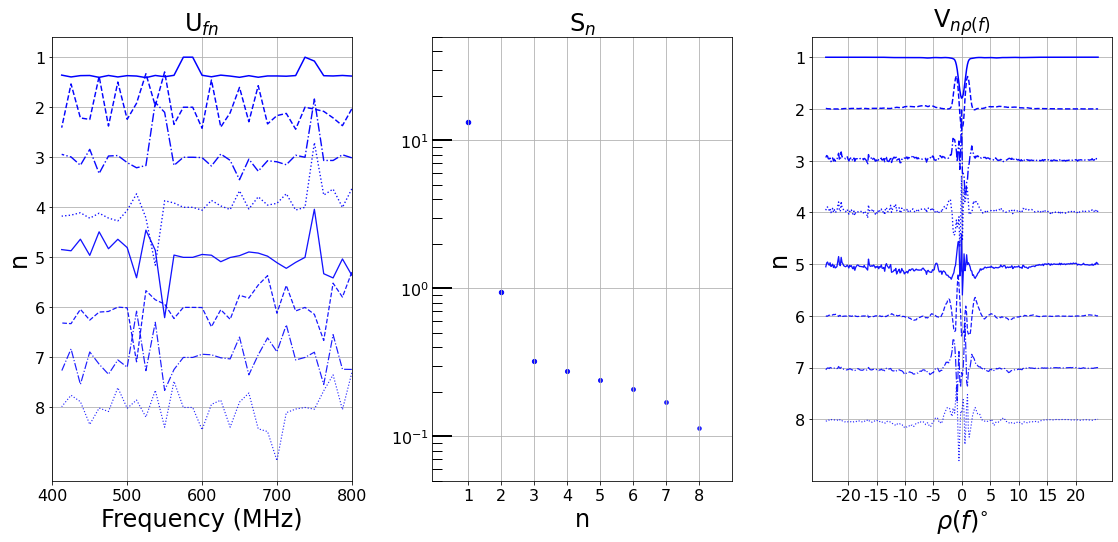}
  \caption{
    The SVD decomposition of the holography data B$_{f,\rho(f)}$ of CygA. We show the first eight modes in the panels. Left: The eigenfunction of frequency (U$_{fn}$). Middle: The eigenvalues (S$_{n}$). Right: The eigenfunction of $\rho(f)$ (V$_{n\rho(f)}$). 
  }
  \label{figure:beam_svd}
\end{figure*}

\begin{figure*}
  \centering
  \includegraphics[width=2\columnwidth]{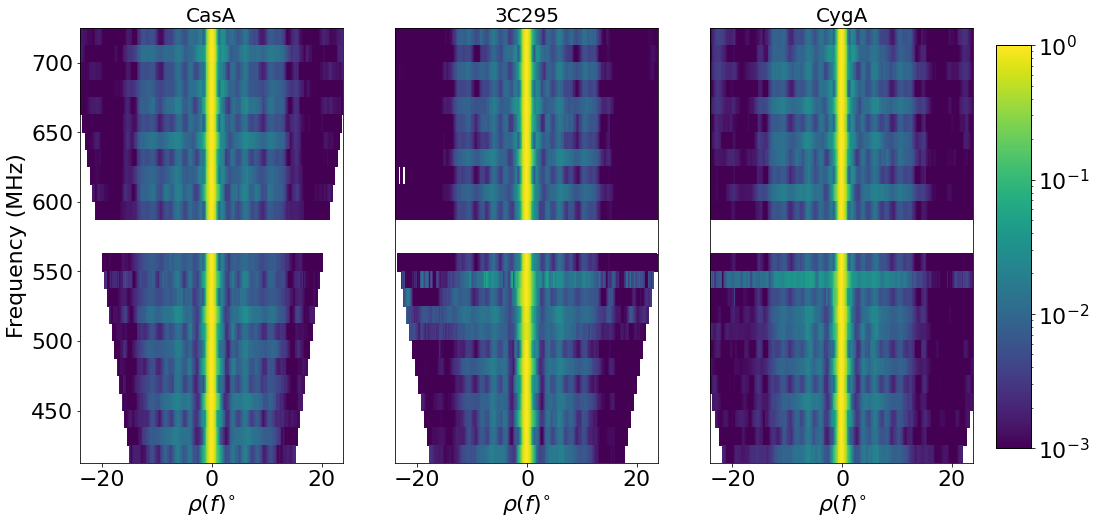}
  \caption{
   The beam-response of three calibrators (CasA, 3C295, and CygA)  in terms of frequency v.s. $\rho(f)$. The frequency spans 400 to 725 MHz with a resolution of 12.5 MHz. $\rho(f)$ spans $-$24$^{\circ}$ to 24$^{\circ}$ in steps of 0.15$^{\circ}$. The RFI channels are masked. For the three calibrators, the holography data do not fully cover the range of $\rho(f)$ in the lower band, and hence there are empty regions which we address with zero-padding.
    }
  \label{figure:beam_panels}
\end{figure*}

\begin{figure}[htp!]
  \centering
  \includegraphics[width=\columnwidth]{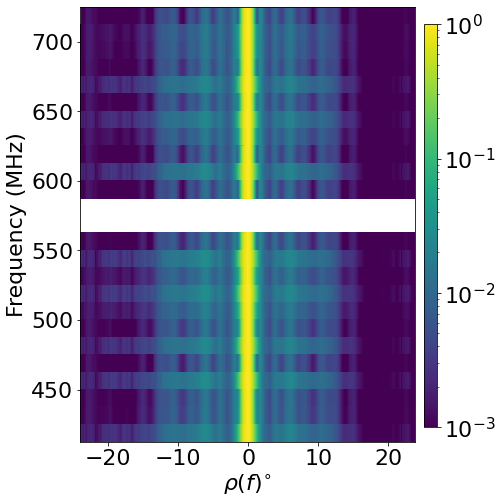}
  \caption{
    The beam model resulting from applying the SVD technique to the holography data of CygA. 
    }
  \label{figure:svd_model_CygA}
\end{figure}

\begin{figure*}[htp!]
  \centering
  \includegraphics[width=2\columnwidth]{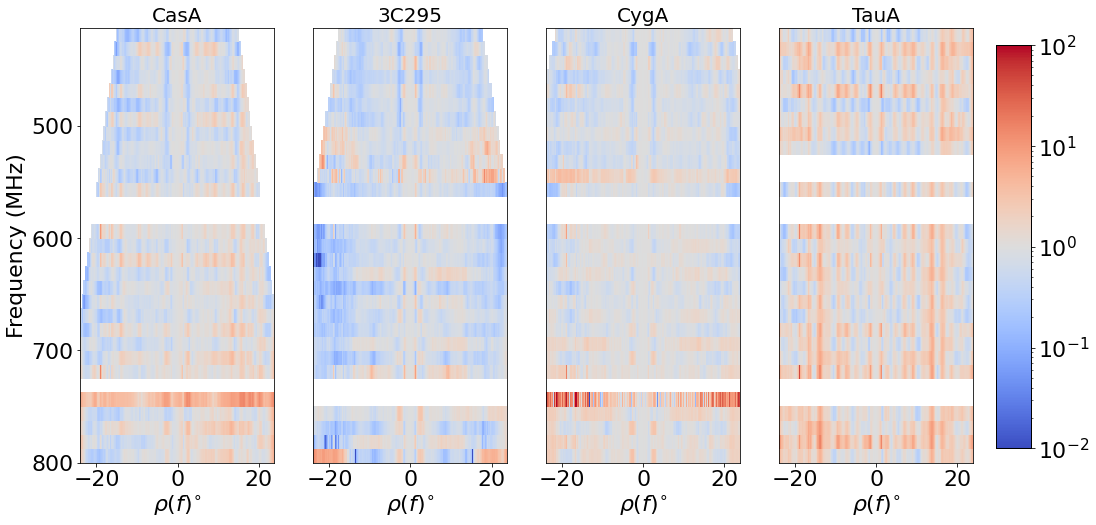}
  \caption{
  Absolute ratio of the holography and beam model for the three calibrators, CasA, 3C295, and CygA. 
  }
  \label{figure:beam_svd_residuals}
\end{figure*}

Appendix Figure \ref{figure:beam_svd} shows the SVD decomposition of the holography data B$_{f,\rho(f)}$ of CygA, from which we use the first two modes to reconstruct the beam model as shown in Figure \ref{table: beam_reponse}. \psedit{In addition, Appendix Figure \ref{figure:beam_panels}, \ref{figure:svd_model_CygA}, and \ref{figure:beam_svd_residuals} shows the holography data, resulting beam model form the SVD decomposition for CygA, and the residuals (the absolute value of data/model) of the three different calibrators.} The residuals are generally around order unity, i.e., the SVD-reconstructed beam model of CygA is
in agreement with the holography data from the other two calibrators
at different declination. Hence, our S/N calibration is independent of the three calibrators.

\section{Determination of the geometric factor, G\label{app: geometric factor}}

\begin{figure}[!htb]
  \centering
  \includegraphics[width=\columnwidth]{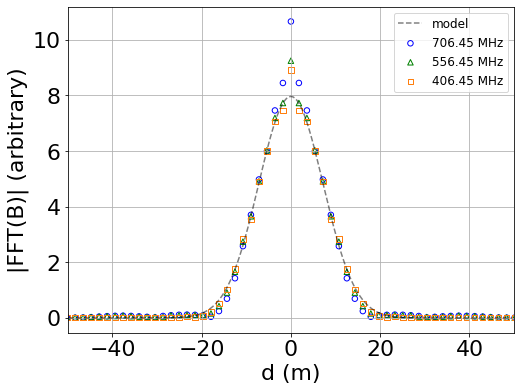}
  \caption{The measurement of the geometrical factor G. The outline markers represent the Fourier transform of the beam response of CygA at different bands, which the beam response is measured per 0.6$^{\circ}$ (a quarter of the resolution of Figure \ref{figure:beam_panels}) for visualization purpose. The dashed line illustrate the Gaussian fitting line, and the FWHM is 16.94 m. 
  }
  \label{figure:beam_spacing}
\end{figure}

The dynamic spectrum of a far side-lobe event shows the spikes \citep[i.e., the ``comb-like'' spectral structure mentioned in][]{2020Natur.587...54C}, which result from the product of the synthesised beam response \citep{2017ursi.confE...4N, 2017arXiv171008591M} and the intrinsic spectral profile across frequencies. The four E-W beams do not fully cover the flux from a far side-lobe event, as summing up the dynamic spectrum of the four E-W beams still shows the spikes. To understand how many E-W beams are required to fully cover the flux of a far side-lobe event, we consider the separation between the beams in the E-W direction as a geometric effect, in which receivers only partially detect the flux from far side-lobe events. Since there are four beams in the E-W direction, we consider the lower bound of the geometric factor G to be 4. To estimate the upper bound of the geometric factor G, we apply a Fourier transform to the frequency-scaled beam response, fit it with a Gaussian profile, and measure the full width at half maximum (FWHM) as 16.94 meters, which is shown in Figure \ref{figure:beam_spacing}. As each cylinder has a width of 20 m and there is a 2 m gap between each cylinder, and therefore the total width of four cylinders is 86 m, we expect that 5 beams in the E-W direction are required to fully cover the flux of a side-lobe event, and therefore the upper bound of the geometric factor G is 5. Hence, we constrain the geometrical factor G to be between 4 and 5. 

